\documentclass[aps,prb,reprint,superscriptaddress]{revtex4-1}
\usepackage{graphicx}% Include figure files
\usepackage{dcolumn}% Align table columns on decimal point
\usepackage{bm}% bold math
\usepackage{amssymb}
\usepackage{epstopdf}

\begin{document}

\preprint{APS/PRB}

\title{Magnetic structures of non-cerium analogues of heavy-fermion Ce$_2$RhIn$_8$: case of Nd$_2$RhIn$_8$, Dy$_2$RhIn$_8$ and Er$_2$RhIn$_8$}

\author{Petr \v{C}erm\'{a}k}
\affiliation{%
J{\"u}lich Centre for Neutron Science JCNS, Forschungszentrum J{\"u}lich GmbH, Outstation at MLZ, Lichtenbergstraße 1, 85747 Garching, Germany
}%
\affiliation{%
 Charles University, Faculty of Mathematics and Physics, Department
of Condensed Matter Physics, Ke~Karlovu~5, 121 16 Prague 2, The
Czech Republic
}%
\email{cermak@mag.mff.cuni.cz}

\author{Pavel Javorsk\'y}
\affiliation{%
 Charles University, Faculty of Mathematics and Physics, Department
of Condensed Matter Physics, Ke~Karlovu~5, 121 16 Prague 2, The
Czech Republic
}%

\author{Marie Kratochv\'\i lov\'{a}}
\affiliation{%
 Charles University, Faculty of Mathematics and Physics, Department
of Condensed Matter Physics, Ke~Karlovu~5, 121 16 Prague 2, The
Czech Republic
}%

\author{Karel Pajskr}
\affiliation{%
 Charles University, Faculty of Mathematics and Physics, Department
of Condensed Matter Physics, Ke~Karlovu~5, 121 16 Prague 2, The
Czech Republic
}%

 \author{Bachir Ouladdiaf}
\affiliation{%
 Institut Laue Langevin, 6 rue Jules Horowitz, BP156, 38042 Grenoble Cedex 9, France
}%

\author{Marie-H\'el\`ene Lem\'ee-Cailleau}
\affiliation{%
 Institut Laue Langevin, 6 rue Jules Horowitz, BP156, 38042 Grenoble Cedex 9, France
}%

\author{Juan Rodriguez-Carvajal}
\affiliation{%
 Institut Laue Langevin, 6 rue Jules Horowitz, BP156, 38042 Grenoble Cedex 9, France
}%

\author{Martin Boehm}
\affiliation{%
 Institut Laue Langevin, 6 rue Jules Horowitz, BP156, 38042 Grenoble Cedex 9, France
}%

\date{\today}

\begin{abstract}
$R_2$RhIn$_8$ compounds (space group P4/mmm, $R$ is a rare-earth element) belong
to a large group of structurally related
tetragonal materials which involves several heavy-fermion
superconductors based on Ce.
%After a short summary of already known magnetic structures from this group and extend the investigation by three new compounds,
We have succeeded to grow single
crystals of compounds with Nd, Dy and Er and following our previous bulk measurements, we performed neutron-diffraction
studies to determine their magnetic structures. The Laue diffraction
experiment showed that the antiferromagnetic order below the N\'eel temperature
is in all three compounds characterized by the propagation vector \textbf{k} = (1/2,
1/2, 1/2). The amplitude and direction of the magnetic moments, as well as
the invariance symmetry of the magnetic structure, were determined by
subsequent experiments using two- and four-circle diffractometers.
The critical exponents were determined from the temperature dependence of
the intensities below $T_N$.
\end{abstract}


\maketitle

\section{Introduction}
The group of  heavy-fermion tetragonal compounds based on the CeIn$_3$ common structural unit became important after the discovery of superconducting state under applied pressure in CeRhIn$_5$ \cite{CeRh115-SC} and later at ambient pressure in CeCoIn$_5$ \cite{CeCo115-SC},  CeIrIn$_5$ \cite{CeIr115-SC} and recently Ce$_2$PdIn$_8$ \cite{CePd218}. This family of structurally related compounds, generally written as Ce$_nT_m$In$_{3n+2m}$ (where $T$ is a transition metal element Co, Rh, Ir, Pd or Pt, $n$ and $m$ are integers), consists of $n$ layers of CeIn$_3$ alternating along the $c$-axis with $m$ layers of $T$In$_2$. The possibility of changing the dimensionality in these materials by varying the $m$ and $n$ together with changing of $T$ element gives scientists a big playground for tuning the ground state properties of these compounds (see Ref. \onlinecite{Ce115-Fisk}).
Since the discovery of similarities between the heavy-fermion superconductivity and the $^3$He magnetic superfluid state it is believed, that these phenomena are mediated by a nearly localized Fermi liquid state and thus, with a magnetic origin \cite{HeAnalogy-Ott1984}.
Hence, a detailed investigation of the magnetic interactions in these materials is of importance to understand their unconventional superconductivity.

The simplest crystal structure in this family of materials forms the cubic CeIn$_3$ ($m$ = 0 and $n$ = 1), where cerium atoms are arranged with a fully 3D character ("13" structure). CeIn$_3$ orders antiferromagnetically (AF) at $T_N$ = 10~K with propagation vector \textbf{k} = (1/2, 1/2, 1/2) \cite{Ce13}. By adding a layer of $T$In$_2$ after every second CeIn$_3$ layer, one can obtain so called "218" structure ($n$ = 2, $m$ = 1), where layers of Ce atoms start to interact quasi two-dimensionally. The only AF ordered cerium compound with the "218" structure is Ce$_2$RhIn$_8$ showing AF transition at $T_N$ = 2.8~K, while other compounds undergo a transition to superconducting state or exhibit a non-fermi liquid behavior. Ce$_2$RhIn$_8$ orders magnetically with the commensurate (C) propagation \textbf{k} = (1/2, 1/2, 0) and a staggered cerium moment of 0.55~$\mu_B$ pointing 52$^\circ$ out of the $ab$-plane \cite{CeRh218-MagStruct}. The stacking of cerium moments within the $ab$-plane remains the same as in CeIn$_3$, but moments stop propagating along the tetragonal $c$-axis.

Adding one layer of $T$In$_2$ between neighborhood cerium planes leads to a complete disappearance of the original cubic cell and to a formation of the so called "115" structure ($m$ = $n$ = 1).
The arrangement of Ce atoms in this type of structure reveals stronger 2D character compared to the 218 structure.
The interest has mainly focused on the 115 compounds in the past years, as they reveal higher superconducting temperatures.
Moreover, their synthesis does not suffer from inclusions and stacking faults as it is often observed in 218 single crystals.
Magnetic order at ambient pressure was found in CeRhIn$_5$ below $T_N$ = 3.8~K. It exhibits similar properties as its 218 analogue, but it forms incommensurate (IC) AF structure propagating with a wave vector \textbf{k} = (1/2, 1/2, 0.297) and cerium magnetic moments of 0.75 $\mu_B$ lying within the $ab$-plane \cite{CeRh115-MagStruct,CeRh115-MagStruct-errata}. The amplitude of the moment constitutes the major part of a value expected from the crystal-field calculations (0.92 $\mu_B$) \cite{CeRh115-CF} which speaks for 4f-localized magnetism. Influence of neighboring Ce layers is decreased leading to IC propagation along the $c$-axis.
The other existing compounds of the cerium 115 family ($T$ = Co, Ir) become superconducting at low temperatures and do not exhibit magnetic order without applied magnetic field. By applying an external magnetic field along the $c$-axis in CeCoIn$_5$ the so called Q-phase appears with magnetic moments of 0.15 $\mu_B$ aligned along the $c$-axis and propagating with the wave-vector \textbf{k} = (0.45, 0.45, 1/2) \cite{CeCo115-MagStruct}. It is questionable whether this magnetic ordering has its origin in the so-called FFLO phase or not, see Ref. \onlinecite{Ce115-Fisk} and references therein. The latest study by Raymond et al. \cite{CeNdCo-115-magstruct} showed the possibility to induce the same Q-phase by a small amount of neodymium doping, raising again the question of the origin of such magnetic ordering.

The recently discovered compound CePt$_2$In$_7$ ("127" structure, $m$ = 2 and $n$ = 1) enhances the 2D character of these compounds: layers of cerium are alternating with two layers of $T$In$_2$. CePt$_2$In$_7$ orders antiferromagnetically below 5.4~K \cite{CePt127}. Coexistence of commensurate $\textbf{k}$ = (1/2, 1/2, 1/4) and incommensurate $\textbf{k}$ = (1/2, 1/2, $\delta$) magnetic structures was revealed by NMR measurement \cite{CePt127-NMR}. The incommensurate component vanishes under pressure and, simultaneously, superconductivity emerges. However the exact magnetic order remains unknown.

In summary, magnetic structures in cerium-based compounds embody a complex behavior resulting from a mixing of competing effects.
To understand magnetic interactions in these compounds, it is useful to follow the evolution of their magnetic structures as a function of different rare-earth elements.
The binary $R$In$_3$ alloys belong to the most studied systems. In contrast to CeIn$_3$,
they all have magnetic ground state with a propagation vector \textbf{k} = (1/2, 1/2, 0). The amplitudes and directions of magnetic moments in the ground state are summarized in Table \ref{mag-structures}. The majority of $R$In$_3$ compounds exhibits a succession of different magnetic phases with decreasing temperature, resulting in a simple commensurate ground state structure. For example the magnetic phase diagram of NdIn$_3$ includes two incommensurate phases in zero magnetic field \cite{Nd13}.

The majority of non-cerium "115" and "218" compounds orders AF and can be split into four groups according to the direction of the easy magnetization axis. Generally, compounds with $R$ = Pr remains paramagnetic (except Pr$_2$PdIn$_8$ \cite{PrPd218}), compounds $R$ = Gd and Sm are nearly isotropic, compounds with $R$ = Nd, Tb, Dy, Ho have the easy magnetization axis along the tetragonal $c$-axis, and the easy magnetization axis lies within the $ab$-plane in the case of compounds with Er and Tm reflecting the crystal-field anisotropy.
As shown in Table \ref{mag-structures}, only a limited number of "218" and "115" magnetic structures has been studied microscopically.
Compounds containing Ga on positions of In atoms form the same structure for heavy rare-earth atoms (Gd-Yb) \cite{ErTmCoGa218}. These intermetallics have similar bulk properties as their indium relatives. Magnetic structures were determined on TbCoGa$_5$ and $R_2$CoGa$_8$ ($R$ = Gd-Tm). All these non-cerium compounds are usually influenced by the RKKY interaction, crystalline electrical field (CEF) effects, and the hybridization between 4f-electrons and conduction electrons \cite{HoCoGa218,GdTbDyCoGa218}.

\begin{table*}[!htbp]
\caption{\label{mag-structures}
Known magnetic structures at ambient pressure and zero magnetic field for $R_nT_m$In$_{3n+2m}$ and $R_nT_m$Ga$_{3n+2m}$ compounds.}
\begin{ruledtabular}
\begin{tabular}{lccddl}
 compound & \textbf{k}-vector & direction\footnotemark[1] &  \multicolumn{1}{c}{amplitude ($\mu_B$)}  & \multicolumn{1}{c}{$T_N$ (K)}    \\
\colrule  %\hline
 \multicolumn{6}{c}{"13"}  \\ \hline
    CeIn$_3$ \cite{Ce13}              & (1/2, 1/2, 1/2) &             & 0.48  & 10    & \footnotemark[2]  \\
    NdIn$_3$ \cite{Nd13}              & (1/2, 1/2, 0)   & $c$-axis    & 2     & 5.9   & \footnotemark[3] \\
    GdIn$_3$ \cite{Gd13}              & (1/2, 1/2, 0)   & $c$-axis    &       & 44    &  \\
    TbIn$_3$ \cite{Tb13-Ho13}         & (1/2, 1/2, 0)   & 10$^\circ$  & 8.4   & 32    &  \\
    DyIn$_3$ \cite{Dy13}              & (1/2, 1/2, 0)   & 27$^\circ$  & 8.8   & 24    &  \\
    HoIn$_3$ \cite{Tb13-Ho13}         & (1/2, 1/2, 0)   & 58$^\circ$  & 9     & 7.9   &  \\
    ErIn$_3$ \cite{Er13}              & (1/2, 1/2, 0)   & [111]     &       & 4.8   &  \\
    TmIn$_3$ \cite{Tm13}              & (1/2, 1/2, 0)   & [111]     & 4.89  & 1.6   & \footnotemark[4] \\
    \hline   \multicolumn{6}{c}{"115"}  \\ \hline
    CeRhIn$_5$ \cite{CeRh115-MagStruct,CeRh115-MagStruct-errata}  & (1/2, 1/2, 0.297) & $ab$-plane & 0.75  & 3.8   &  \\
    CeCoIn$_5$ \cite{CeCo115-MagStruct}  & (0.44, 0.44, 1/2) & $c$-axis & 0.15  & 0.3   & \footnotemark[5] \\
    Ce$_{0.95}$Nd$_{0.05}$CoIn$_5$ \cite{CeNdCo-115-magstruct}
                        & (0.45, 0.45, 1/2)             &             &       & 0.9   &  \\
    NdRhIn$_5$ \cite{NdRh115}         & (1/2, 0, 1/2)   & $c$-axis    & 2.5   & 11    &  \\
    GdRhIn$_5$ \cite{GdRh115}         & (1/2, 0, 1/2)   & $b$-axis    &       & 39    &  \\
    TbRhIn$_5$ \cite{TbRh115}         & (1/2, 0, 1/2)   & $c$-axis    & 9.5   & 47.3  &  \\
    DyRhIn$_5$ \cite{RRh115-thesis}   & (1/2, 0, 1/2)   & $c$-axis    & 8.1   & 28.1  &  \\
    HoRhIn$_5$ \cite{RRh115-thesis}   & (1/2, 0, 1/2)   & $c$-axis    & 7.6   & 15.8  &  \\
    TbCoGa$_5$ \cite{TbCoGa115}       & (1/2, 0, 1/2)   & $c$-axis    &       & 36.2  &  \footnotemark[6] \\
    HoCoGa$_5$ \cite{HoCoGa115}       & (1/2, 0, 1/2)   & $c$-axis    &       & 9.7   &  \footnotemark[7] \\
    \hline   \multicolumn{6}{c}{"218"}  \\ \hline
    Ce$_2$RhIn$_8$ \cite{CeRh218-MagStruct}         & (1/2, 1/2, 0) & 38$^\circ$ \footnotemark[8]  &  0.55  & 2.8  &  \\
    Tb$_2$RhIn$_8$ \cite{TbRh218}     & (1/2, 1/2, 1/2) &            \footnotemark[9] &       & 42.8  &   \\
    Gd$_2$IrIn$_8$ \cite{GdIr218}     & (1/2, 0, 0)     & $ab$-plane \footnotemark[8] &       & 40.8  &   \\
    Sm$_2$IrIn$_8$ \cite{SmIr218}     & (1/2, 0, 0)     & $ab$-plane \footnotemark[8] &       & 14.2  &   \footnotemark[10] \\
    Gd$_2$CoGa$_8$ \cite{GdTbDyCoGa218} & (1/2, 1/2, 1/2) & $ab$-plane \footnotemark[11] &       & 20.0 & \\
    Tb$_2$CoGa$_8$ \cite{GdTbDyCoGa218} & (1/2, 1/2, 1/2) & $c$-axis   \footnotemark[11] &       & 28.5 & \\
    Dy$_2$CoGa$_8$ \cite{GdTbDyCoGa218} & (1/2, 1/2, 1/2) & $c$-axis   \footnotemark[11] &       & 15.2 & \\
    Ho$_2$CoGa$_8$ \cite{HoCoGa218}     & (1/2, 1/2, 1/2) & $c$-axis                     &       & 5.1  & \\
    Er$_2$CoGa$_8$ \cite{ErTmCoGa218}   & (0, 1/2, 0)     & $a$-axis   \footnotemark[8]  & 4.71  & 3.0  & \\
    Tm$_2$CoGa$_8$ \cite{ErTmCoGa218}   & (1/2, 0, 1/2)   & $a$-axis   \footnotemark[11] & 2.35  & 2.0  & \\
\end{tabular}
\end{ruledtabular}
\footnotetext[1]{Value in degrees means inclination from the $c$-axis.}
\footnotetext[2]{Magnetic moment direction cannot be determined by neutron diffraction.}
\footnotetext[3]{This C structure is stabilized below 4.7 K. Above this temperature there is a mixture of IC phases.}
\footnotetext[4]{Compound also contains \textbf{k} = (0, 0, 1/2) propagation and an IC component.}
\footnotetext[5]{In the magnetic field 11~T applied along the [1-10] direction.}
\footnotetext[6]{Magnetic structure for phase between 5.4 and 36.2~K.}
\footnotetext[7]{This C structure is stabilized below 7.4 K. Between this temperature and $T_N$ exists an IC phase with $k$= (0.5, 0, 0.359).}
\footnotetext[8]{$+-+-$ stacking along the $c$-axis.}
\footnotetext[9]{$+--+$ stacking along the $c$-axis.}
\footnotetext[10]{Direction of the moments was determined to be 18$^\circ$ from the $a$-axis.}
\footnotetext[11]{$++--$ stacking along the $c$-axis.}

\end{table*}

In this work we report the determination of the magnetic structures of
$R_2$RhIn$_8$ (with $R$ = Nd, Dy, and Er) compounds using the single crystal neutron diffraction technique.
As the Pr compound from this family exhibits a non-magnetic
singlet ground state\cite{Pr}, the Nd-based compound is the natural
candidate that should be primarily investigated.
Moving along the lanthanide series, we have chosen Dy and Er based compounds for a detailed study.
Dy$_2$RhIn$_8$ represents a typical member of heavy rare-earth compounds, having the same direction of the easy magnetization axis as Nd$_2$RhIn$_8$ \cite{NdRh115-phase} but a much larger amplitude of the ordered magnetic moments \cite{RRh218-our}. Both neodymium and dysprosium based compounds exhibit very sharp steps in magnetization curves indicating the existence of a field-induced phase in their magnetic phase diagram \cite{RRh218-our}. A similar phase diagrams were reported for their 115 relatives \cite{RRh115-CF2006}, cobalt-galium relatives \cite{RCoGa218-bulk} and Tb$_2$RhIn$_8$ \cite{RRh218-our}, pointing to a similar magnetic scenario.
On the other hand, Er$_2$RhIn$_8$ represents a compound where the easy magnetization axis lies within the $ab$-plane. Rather smooth steps in the magnetization curves were observed in this case when applying a field along the twofold [110] direction \cite{RRh218-our}. In total, bulk properties of all three studied compounds are similar to the structurally related $R$RhIn$_5$ \cite{RRh115-CF2007} and $R_2$CoGa$_8$ \cite{RCoGa218-bulk} compounds.

In order to determine the magnetic structure in these materials, we have performed two types of neutron diffraction
experiments. First, neutron Laue diffraction images were taken to explore the reciprocal space and find the propagation vectors. Subsequently standard
two- or four-circle diffraction experiments were carried out to determine the magnetic structures in detail.

\section{Experiment}

Single crystals of Nd$_2$RhIn$_8$, Dy$_2$RhIn$_8$ and Er$_2$RhIn$_8$ were prepared by the solution
growth method from an indium flux \cite{growth}. The elements with starting
compositions 2:1:40, 2:1:30 and 2:1:50, respectively, were put into alumina
crucibles, sealed under high vacuum and heated up to 910~$^{\rm{o}}$C.
The mixture was then slowly cooled down to 400~$^{\rm{o}}$C
where the remaining indium solution was centrifuged.
In this way we obtained plate-shaped cuboid single crystals.
The samples selected for further macroscopic and microscopic measurements were
of sizes about 1.7x1.5x0.8~mm$^3$, 2.5x0.4x0.4~mm$^3$ and 4x1.5x0.2~mm$^3$, respectively. The $c$-axis was always oriented
perpendicular to the plate. The chemical composition and homogeneity were verified by an energy-dispersive X-ray
detector Bruker AXS and the tetragonal space group \emph{P4/mmm} together with lattice parameters were confirmed
on single crystal X-ray RIGAKU RAPID II diffractometer. Atomic positions of
$R_2$RhIn$_8$ are presented in the Table \ref{atomic-positions}.

\begin{table}[b]
\caption{\label{atomic-positions}
Atomic Positions of $R_2$RhIn$_8$, space group \emph{P4/mmm}}
\begin{ruledtabular}
\begin{tabular}{lllll}
  atom &   & $x$ & $y$ & $z$ \\
\colrule  %\hline
$R$ & 2g & 0   & 0   & $z$($R$)\\
Rh  & 1b & 0   & 0   & 1/2\\
In1 & 2f & 1/2 & 0   & 0\\
In2 & 4i & 1/2 & 0   & $z$(In2)\\
In3 & 2h & 1/2 & 1/2 & $z$(In3)\\
\end{tabular}
\end{ruledtabular}
\end{table}

Neutron Laue diffraction experiments on Nd$_2$RhIn$_8$ and Dy$_2$RhIn$_8$ crystals
were performed on the VIVALDI instrument at Institute Laue Langevin (ILL), Grenoble
\cite{VIVALDI}. The Laue patterns were recorded in the paramagnetic
state at 30 and 40~K, respectively, and in the ordered state at 2~K.
In order to maximize number of observed reflections and to discover any
possible purely magnetic intensities, the crystal was mounted with
obvious symmetry axes well away from the vertical axis. Nine
patterns at 10 degree intervals of rotation about the vertical
axis were taken at each temperature. Each pattern was exposed for 115
minutes.
The propagation vector of the Er$_2$RhIn$_8$ was determined on the CYCLOPS instrument also at ILL \cite{CYCLOPS}.
Laue patterns were collected in the paramagnetic (8~K) and in the antiferromagnetic (1.5~K) state.
26 patterns (each exposed for 15 minutes) at 5 degree intervals were taken at each temperature.
A series of Laue patterns with changing temperature were also taken in order to reveal possible
phase transitions.
All Laue patterns, both from the VIVALDI and the CYCLOPS instruments,
were indexed and integrated using the Esmeralda Laue Suite software \cite{ESMERALDA}.

The four-circle neutron diffraction experiments were
performed for the Nd and Dy samples on the
D10 diffractometer at ILL, with a
wavelength $\lambda = 2.36$~$\textrm{\AA}$ using pyrolytic graphite monochromator and filter before the sample. The reflections were measured as
$\omega$-scans. After cooling the samples to 2~K, cell parameters
and orientation were refined on the basis of 41 (Nd) and 20 (Dy)
strong nuclear reflections using the program RAFD9 \cite{rafd9}. Then a set of
reflections at 2~K and temperature dependencies of selected magnetic and nuclear reflections were measured.
All reflections were integrated and corrected for Lorentz factor using the program RACER \cite{racer}.

In the case of Er$_2$RhIn$_8$, we have used the triple axis spectrometer IN3 at ILL.
We measured the reflections in the elastic condition at $\lambda = 2.36$~$\textrm{\AA}$
using $\omega$-scans as well.
The sample was mounted with the [110] and the [001] lattice vectors in the scattering plane.
After cooling to 1.5~K, the tilt of the sample was adjusted by a goniometer and lattice parameters were refined.
Contrary to D10, IN3 has only $^3$He detector tube.
All measured datasets were fitted with Gaussian profiles
and the integrated intensities were corrected for the Lorentz factor.

Moreover, the integrated intensities of all reflections were corrected for absorption in the crystal using
the program DATAP \cite{datap}.
Used absorption coefficients together with the number of measured reflections are listed in Table \ref{D10-arrangement}.
The obtained raw data were reduced using the program DataRed \cite{Fullprof}.
The program FullProf \cite{Fullprof} was used for the refinement of the nuclear and
magnetic structures. The extinction correction was refined using the
Zachariasen formula \cite{extinkce} with anisotropic correction
(Ext-Model=4 in FullProf software).

\begin{table}[b]
\caption{\label{D10-arrangement}
Summary of performed single crystal diffraction experiments}
\begin{ruledtabular}
\begin{tabular}{lccc}
  & Nd$_2$RhIn$_8$ & Dy$_2$RhIn$_8$ & Er$_2$RhIn$_8$ \\
\colrule  %\hline
    instrument & D10 & D10 & IN3 \\
    absorbtion coefficient (cm$^{-1}$) & 9.018 & 19.134 & 10.707 \\
   \multicolumn{4}{l}{number of measured reflections (nonequivalent) } \\
    nuclear  & 364 (70) & 350 (68) & 60 (21) \\
    magnetic & 461 (50) & 383 (38) & 57 (15) \\

\end{tabular}
\end{ruledtabular}
\end{table}

\section{Results and Discussion}

\begin{figure}[b]
\centering
\resizebox{0.5\textwidth}{!}{%
    \includegraphics[scale=0.6]{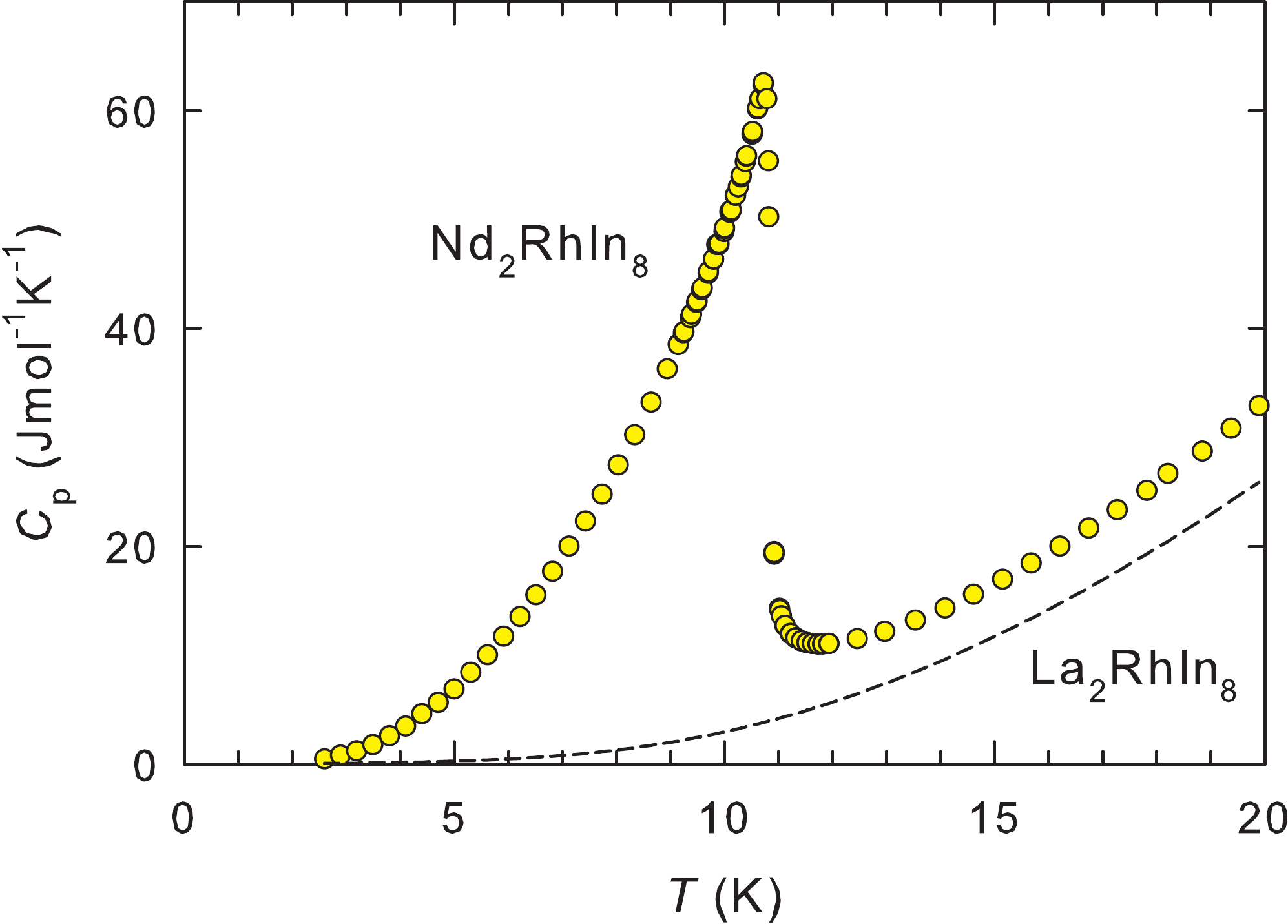}
}
\vspace{0mm} \caption{\label{Cp} Specific heat of
Nd$_2$RhIn$_8$. The dashed line shows the specific heat of a
non-magnetic analogue La$_2$RhIn$_8$. \cite{nonmag} }
\end{figure}

The specific heat of Nd$_2$RhIn$_8$ was measured to compare the
magnetic characteristics of our sample with previously
published data \cite{NdRh115-phase}. The $C_p$ vs $T$ dependence (see Fig.\ref{Cp})
shows a well pronounced $\lambda$-type anomaly corresponding to
the magnetic phase transition. The ordering temperature $T_N =
(10.8 \pm 0.1)$~K can be deduced from our data, in a good
agreement with the previously published value of $T_N =
10.7$~K \cite{Pagliuso2000}. Also the measured absolute values
and the magnetic entropy (not shown here), determined after
subtraction of specific heat of La$_2$RhIn$_8$ taken from Ref.
\onlinecite{nonmag}, correspond well to the values reported by
Pagliuso \cite{Pagliuso2000}. As the specific heat does not show any
sign of a further phase transition down to 2 K, we expect a
single magnetic phase in zero magnetic field.
Similar conclusions can be made for dysprosium and erbium
compounds, based on our previous measurements performed on the
same piece of single crystals \cite{RRh218-our}.

\begin{figure}[h!]
\centering
\resizebox{0.5\textwidth}{!}{%
    \includegraphics{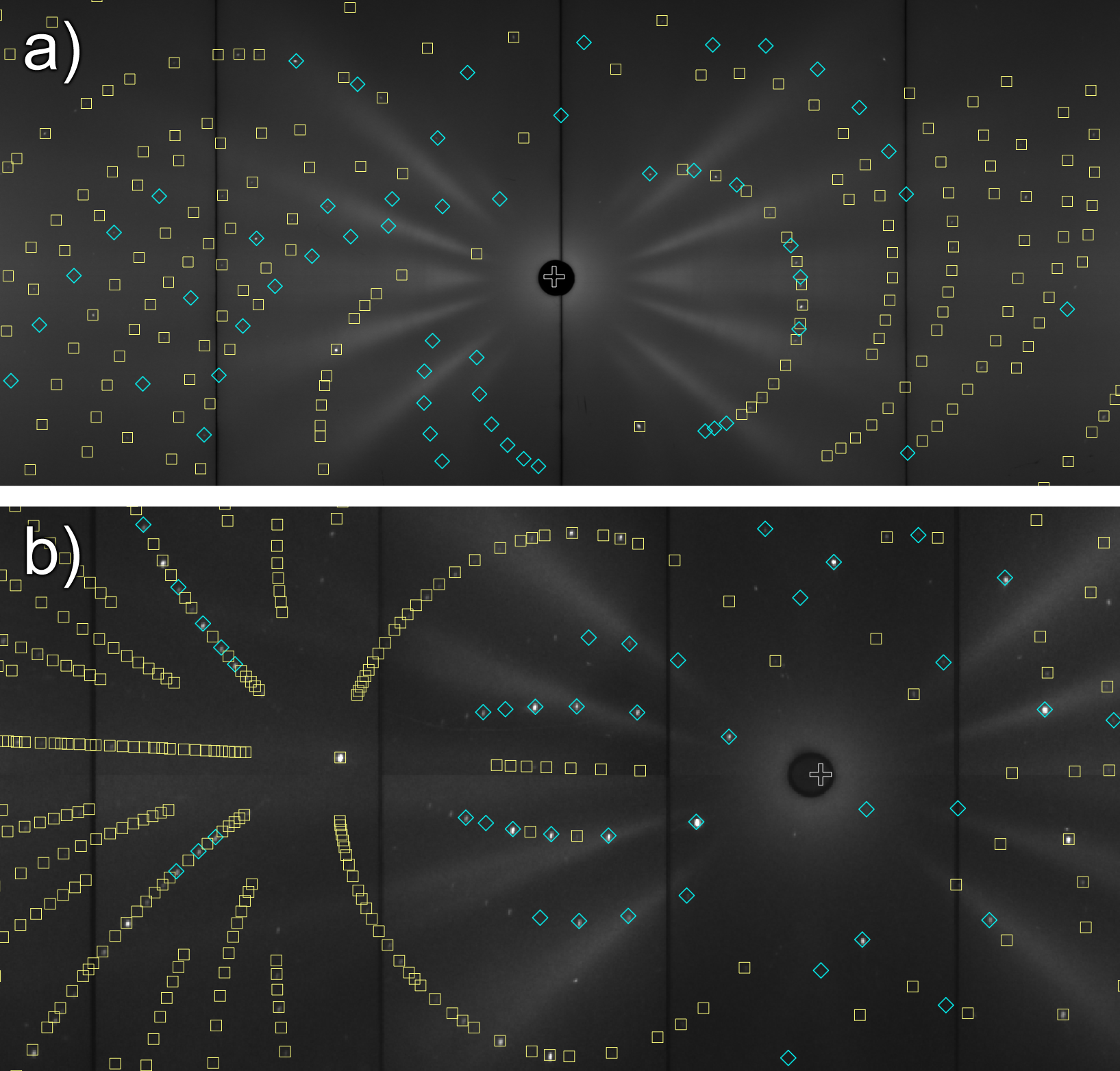}
}
\vspace{0mm} \caption{\label{Laue} Laue picture of
a) Nd$_2$RhIn$_8$, b) Er$_2$RhIn$_8$ taken at 2 and 1.5 K, respectively.
Yellow squares denote nuclear reflections while blue diamonds denote magnetic ones. The diffuse streaks
correspond to the textured powder pattern due to Al in the cryostat. }
\end{figure}

\begin{figure}[h!]
\centering
\resizebox{0.5\textwidth}{!}{%
    \includegraphics{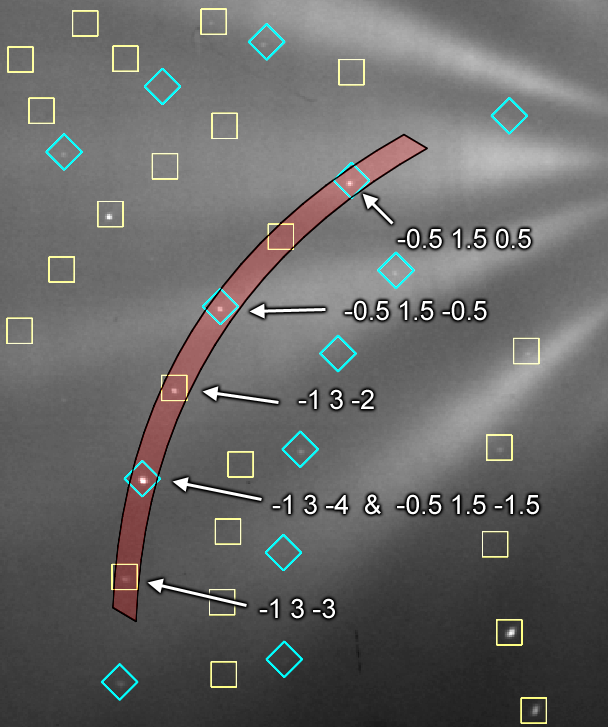}
}
\vspace{0mm} \caption{\label{Lauecut} Part of
the Laue picture of Nd$_2$RhIn$_8$ with diffraction spots marked as on Fig. \ref{Laue}.
Red color denotes area of integration along the reciprocal  [$hkl$]* (with $h$=-1, $k$=3 and $l$=$n$)
direction which correspond to the intensities shown in Fig. \ref{LaueInt}.}
\end{figure}

\begin{figure}[h!]
\centering
\resizebox{0.5\textwidth}{!}{%
    \includegraphics{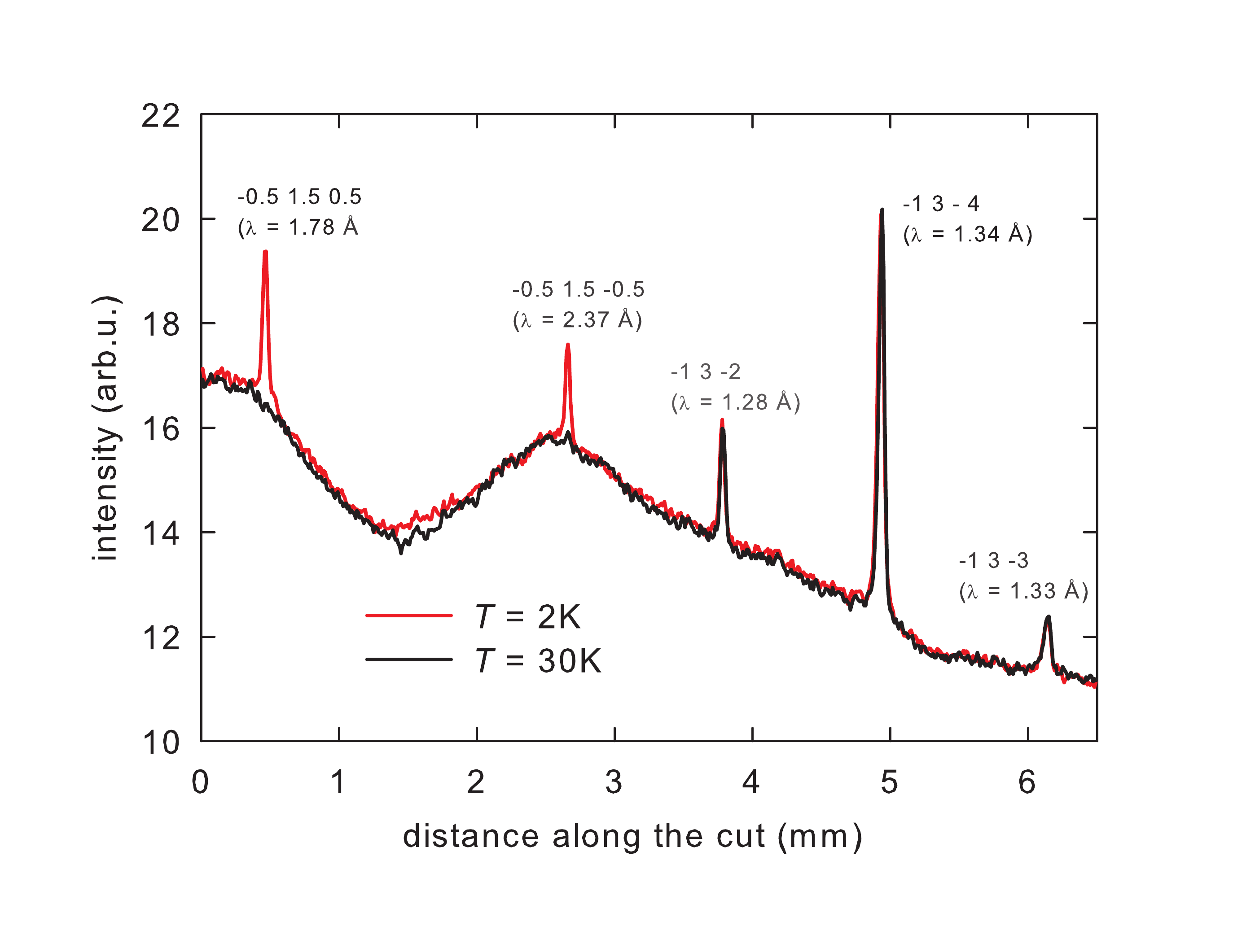}
}
\vspace{0mm} \caption{\label{LaueInt} Diffraction
intensities taken from a cut through a Laue picture of Nd$_2$RhIn$_8$ as indicated
in Fig. \ref{Lauecut}b. Note that individual positions could
correspond simultaneously to several reflections
that are overlapped with different wavelengths.}
\end{figure}

The overall Laue patterns for Nd$_2$RhIn$_8$ and Er$_2$RhIn$_8$
are represented in Fig. \ref{Laue}.
All the observed diffraction spots at paramagnetic temperature
can be indexed assuming the tetragonal structure with the space group \emph{P4/mmm}.
At the cryostat base temperature, a large number of new,
purely magnetic reflections, appear. All magnetic reflections in all three compounds can be
described by a single propagation vector $\textbf{k}$ = (1/2, 1/2, 1/2).
To illustrate this observation,
we show a smaller cut of the Laue picture of Nd$_2$RhIn$_8$ in Fig.\ref{Lauecut}.
The intensities along the [001] crystallographic direction,
indicated in Fig.\ref{Lauecut} b), are then shown in Fig.\ref{LaueInt}.
The knowledge of the propagation vector was subsequently used during
the further single crystal diffraction experiments.

\begin{figure}[h!]
\centering
\resizebox{0.5\textwidth}{!}{%
    \includegraphics{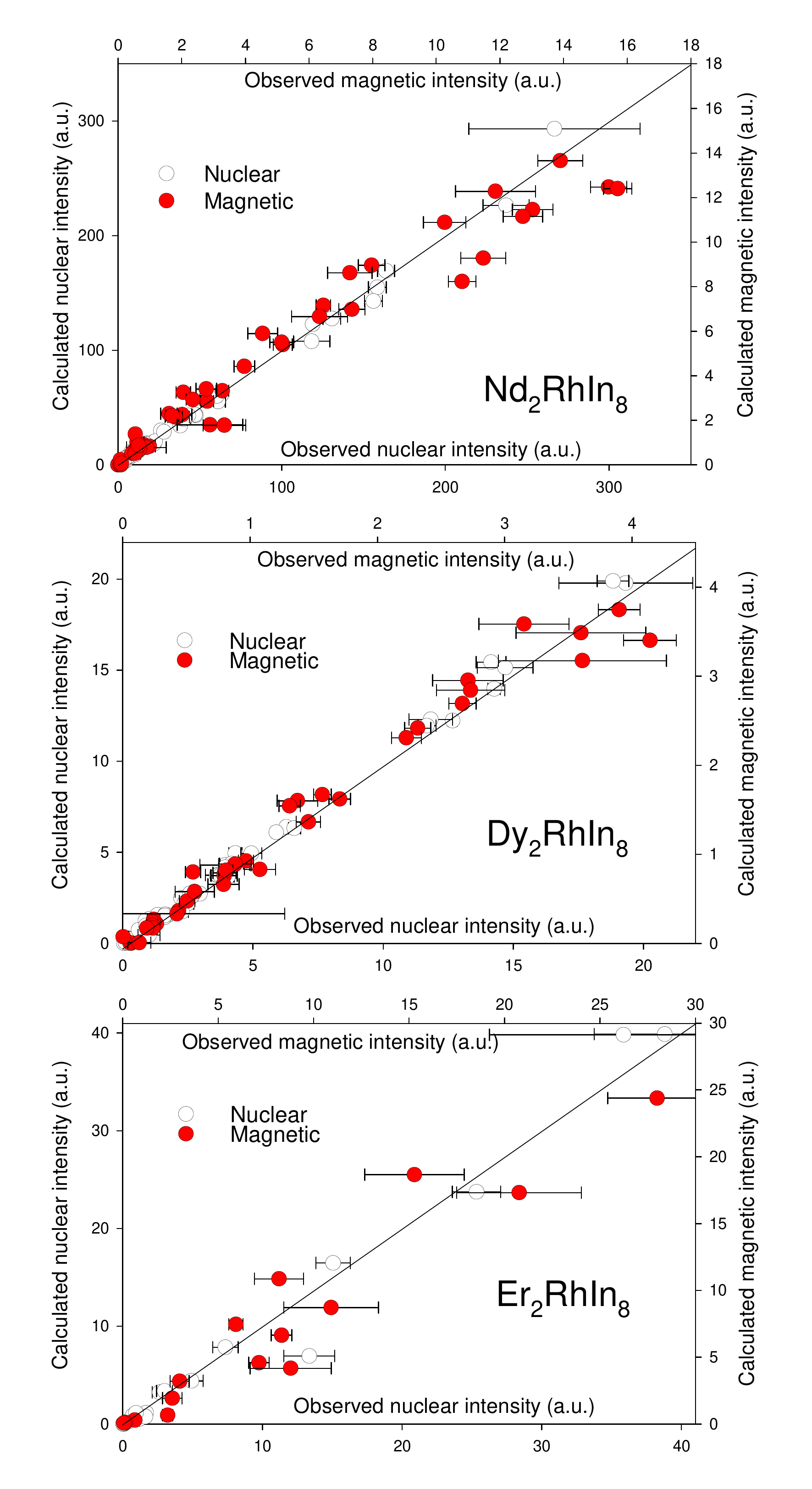}
}
\vspace{0mm} \caption{\label{Intensity} Observed
and calculated integrated nuclear and magnetic intensities. The calculated
intensities correspond to the parameters given in Table \ref{parameters}.
Arbitrary units are used, but values for each compound are scaled together
with the same ratio.}
\end{figure}

The structural parameters at the lowest temperature are summarized in Table \ref{parameters} and
the observed vs. calculated integrated nuclear intensities are depicted in Fig. \ref{Intensity}.

\begin{table}
\caption{Structural and magnetic parameters of $R_2$RhIn$_8$ at
$T$=2~K.} \label{parameters}
    \begin{tabular}{ l l l l }
    \hline
     $R$       & Nd        & Dy       & Er      \\
    \hline
    \multicolumn{4}{l}{lattice parameters}      \\
    \hline
     a ($\textrm{\AA}$) & 4.6213(9) &  4.572(2) & 4.552(2)  \\
     c ($\textrm{\AA}$) & 12.113(3) &  11.96(1) & 11.980(2) \\
    \hline
    \multicolumn{4}{l}{atomic positions along the $c$-axis}     \\
    \hline
     $R$     &  0.3083(3) & 0.3095(2) &  0.311(1)  \\
     In(2)   &  0.3059(6) & 0.3078(7) &  0.311(1)  \\
     In(3)   &  0.1212(4) & 0.1226(5) &  0.125(2)  \\
    \hline
    \multicolumn{4}{l}{magnetic structure }        \\
    \hline
     $\textbf{k}$    & \multicolumn{3}{c}{(1/2, 1/2, 1/2)}  \\
     $\mu$ ($\mu_B$) & 2.53(9)   & 6.9(3)    & 6.4(1.4)   \\
     direction       & $c$-axis  & $c$-axis  & $ab$-plane \\
     $c$-stacking    & $++--$    & $++--$    & $++--$     \\
     $T_N$ (K)       & 10.63(4)  & 24.24(8)  & 3.70(6)    \\
     $\beta$         & 0.22(3)   & 0.20(1)   & 0.16(2)    \\
    \hline
    \multicolumn{4}{l}{reliability factors }  \\
    \hline
     nuclear  $RF^2 $  & 6.70  & 5.58 & 11.8  \\
     nuclear  $RF$     & 5.34  & 4.40 & 9.94  \\
     nuclear  $\chi^2$ & 3.14  & 2.58 & 3.57  \\
     magnetic $RF^2 $  & 15.5  & 9.46 & 20.2  \\
     magnetic $RF$     & 9.83  & 6.95 & 13.2  \\
     magnetic $\chi^2$ & 6.01  & 2.65 & 8.29  \\
    \end{tabular}
\end{table}

\begin{figure}[h!]
\centering
\resizebox{0.5\textwidth}{!}{%
    \includegraphics{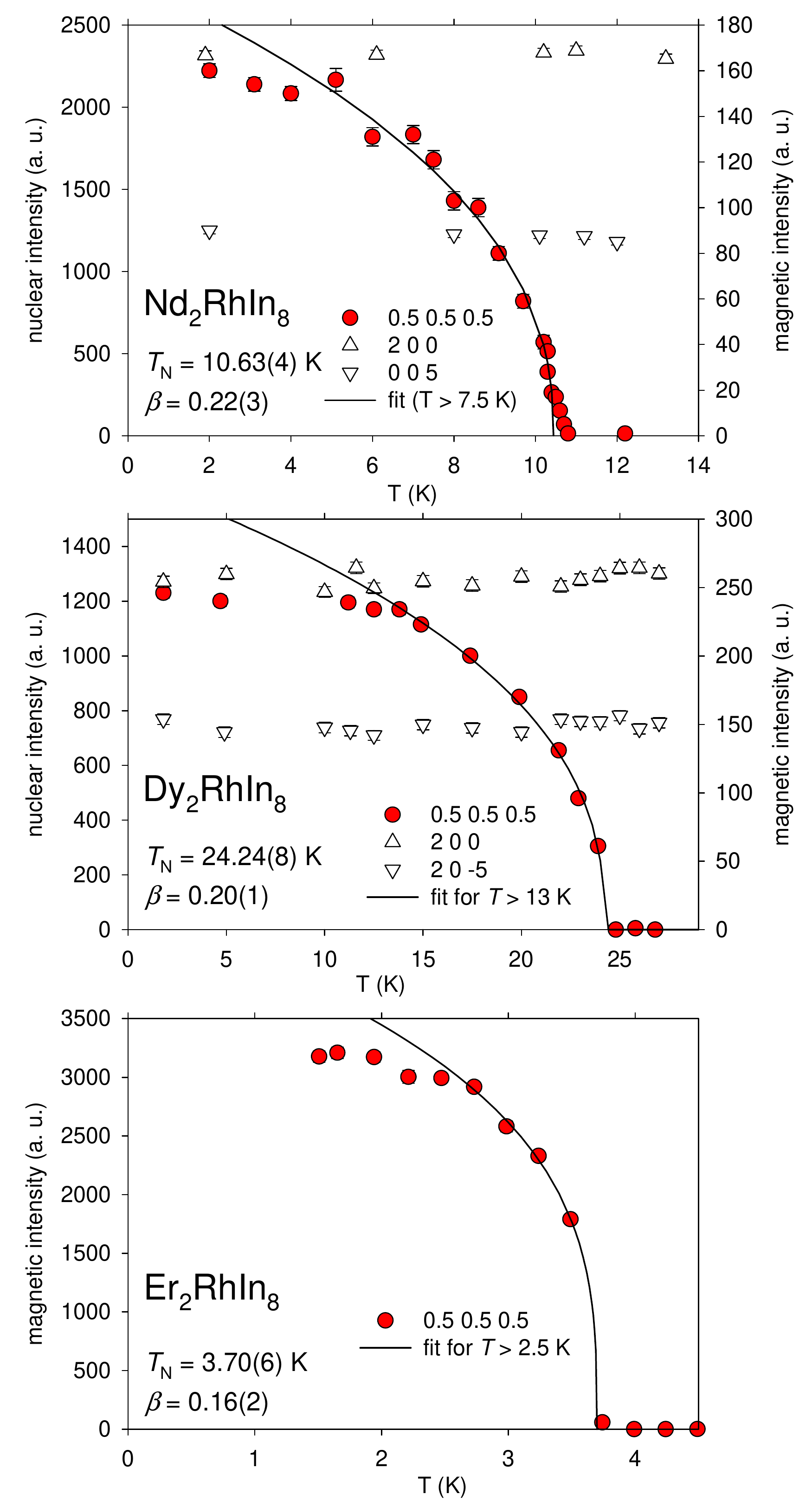}
}
\vspace{0mm} \caption{\label{Tdep} Temperature
dependence of intensities of selected reflections. The full line
is a fit to the equation \ref{crit}. }
\end{figure}

The temperature dependence of selected nuclear intensities of
Nd$_2$RhIn$_8$ and Dy$_2$RhIn$_8$ is shown in Fig.\ref{Tdep}.
We observed no change in intensity above and below the transition
temperature, indicating that there is no contribution with
$\textbf{k}$ = (0, 0, 0) propagation vector.
Similar conclusion can be deduced from the temperature
dependence of Laue patterns from CYCLOPS (not shown) for Er$_2$RhIn$_8$,
where no change in nuclear intensities is observed as well.

In order to restrict the number of possible magnetic structures,
we applied symmetry arguments as developed in the representation
analysis \cite{RepAnaBertaut}. The different irreducible representations
with their associated basis vectors have been calculated with
the help of the BasIreps program \cite{Fullprof} using the
previously measured propagation vector $\textbf{k}$ = (1/2, 1/2, 1/2).
The little group (or group of the propagation vector)
coincides with the space group G$_\textbf{k}$=\emph{P4/mmm} (all rotational symmetry operators of \emph{P4/mmm} leave invariant the propagation vector),
so the small representations coincides with
the full irreducible representations of the space group.
There are together 10 irreducible
representations (\emph{irreps}) associated with the $\textbf{k}$ = (1/2, 1/2, 1/2) propagation vector.
Two of them, $\Gamma_9$ and $\Gamma_{10}$, are
two-dimensional and remaining 8 are one-dimensional. However,
the global reducible magnetic representation of the $R$ 2g-site can be
decomposed in \emph{irreps} as
%% 2g site with the point group $\mathrm{C_{4v}}$ (4mm)
$\Gamma_{2g} = \Gamma_2 + \Gamma_7 + \Gamma_9 + \Gamma_{10}$.
Because there are always two magnetic sublattices corresponding to the 2g Wyckoff site within the unit cell,
the basis vectors have six components each.
The first three correspond to the magnetic moment
components of the $R$ atom at the position with $x$,$y$,$z$ site symmetry (R1) and
the other three to those of the atom at the $-x$,$y$,$-z+1$ site (R2). By
making linear combinations of the basis vectors within the same
irreducible representation we obtain the
vectors representing the components of the magnetic moments of both atoms.
These combinations are summarized in Table \ref{representation}. One can
see, that in the case of the one-dimensional representations $\Gamma_2$ and
$\Gamma_7$ there is only a single free parameter \emph{u} describing the
magnetic structure. For the two-dimensional representations $\Gamma_9$
and $\Gamma_{10}$, there are, in general, two parameters \emph{u} and \emph{v}.
In both cases the difference between $\Gamma_2$ and $\Gamma_7$, or
$\Gamma_9$ and $\Gamma_{10}$, respectively, resides in the either
parallel, or antiparallel coupling between the two rare-earth sublattices.
As the propagation vector is
$\textbf{k}$ = (1/2, 1/2, 1/2), the magnetic unit cell is doubled in $x$,$y$,$z$ direction and the direction of the moments in the neighboring (chemical) unit cells have to be opposite.

\begin{table}
\caption{Direction of magnetic moments for all possible irreducible
representations corresponding to the propagation wave vector $\textbf{k}$ =
(1/2, 1/2, 1/2) and the magnetic 2g site in the \em{P4/mmm} space group.}
\label{representation}
    \begin{tabular}{ l l l l l }
    \hline
        site & $\Gamma_2$  & $\Gamma_7$ & $\Gamma_9$ & $\Gamma_{10}$  \\ \hline
        R1 \qquad \qquad &  0 0 \emph{u} \qquad \qquad & 0 0 \emph{u} \qquad \qquad & \emph{u -v} 0 \qquad \qquad & \emph{u v} 0 \\
        R2 \qquad \qquad &  0 0 \emph{-u} \qquad \qquad & 0 0 \emph{u} \qquad \qquad & \emph{-u v} 0 \qquad \qquad & \emph{u v} 0 \\
     \hline
    \end{tabular}
\end{table}

From the point of view of the invariance symmetry of the spin configurations described by the above representations, it is easy to
determine the Shubnikov groups for all of them using tools like the Bilbao Crystallographic Server \cite{BCS1,BCS2} or the suite of program existing in the
ISOTROPY site \cite{ISOTROPY}. In the case of 1D \emph{irreps} $\Gamma_2$ and $\Gamma_7$ the Shubnikov group is tetragonal and the same: \emph{I$_c$4/mcm} in BNS notation, or
\emph{P$_I$4/mm'm'}in OG notation, except that the $z$-coordinate of the R1 atom is different for the two \emph{irreps}. For the
2D \emph{irreps} we have more possibilities because we can select different directions in the representation space. These directions correspond
to particular values of \emph{u} and \emph{v}. The direction (\emph{a},0) corresponds to \emph{v}=0 and the direction
(0,\emph{b}) corresponds to \emph{u}=0 and all of them gives rise to the same orthorhombic Shubnikov group for both 2D \emph{irreps} $\Gamma_9$ and $\Gamma_{10}$:
\emph{F$_S$mmm} in BNS notation, or \emph{P$_I$mmm} in OG notation, with different atom positions for each representation and directions.
For the direction (\emph{a},\emph{a}) we have \emph{u}=\emph{v} and the symmetry is also orthorhombic: \emph{I$_b$mma} in BNS notation,
or \emph{C$_I$m'mm} in OG notation for both 2D \emph{irreps}. A general direction in the representation space correspond to different values of
\emph{u} and \emph{v} and lowers the symmetry to monoclinic: \emph{C$_a$2/m} in BNS notation or \emph{P$_C$2/m} in OG notation for both 2D \emph{irreps}.
The differences for each \emph{irrep} correspond always to different positions of the magnetic atoms in the standard setting of the Shubnikov group.
The combination of two \emph{irreps}, like $\Gamma_2 + \Gamma_9$, lowers still the symmetry to triclinic in the general case.

A good agreement between observed and calculated intensities of Nd$_2$RhIn$_8$ and Dy$_2$RhIn$_8$ is obtained for
magnetic moments pointing along the $c$-axis with their parallel alignment within one unit cell and corresponding to $\Gamma_2$.
For Er$_2$RhIn$_8$ the fitting procedure showed that the far best agreement is obtained with the model
$\Gamma_{10}$ where the magnetic moments in the unit cell lie in the $ab$-plane pointing the same direction.
For Er$_2$RhIn$_8$ only reflections within the (-110) scattering plane could be measured, which were not sufficient to determine the exact direction of the moments within the $ab$-plane.

\begin{figure}[h!]
\centering
\resizebox{0.5\textwidth}{!}{%
    \includegraphics{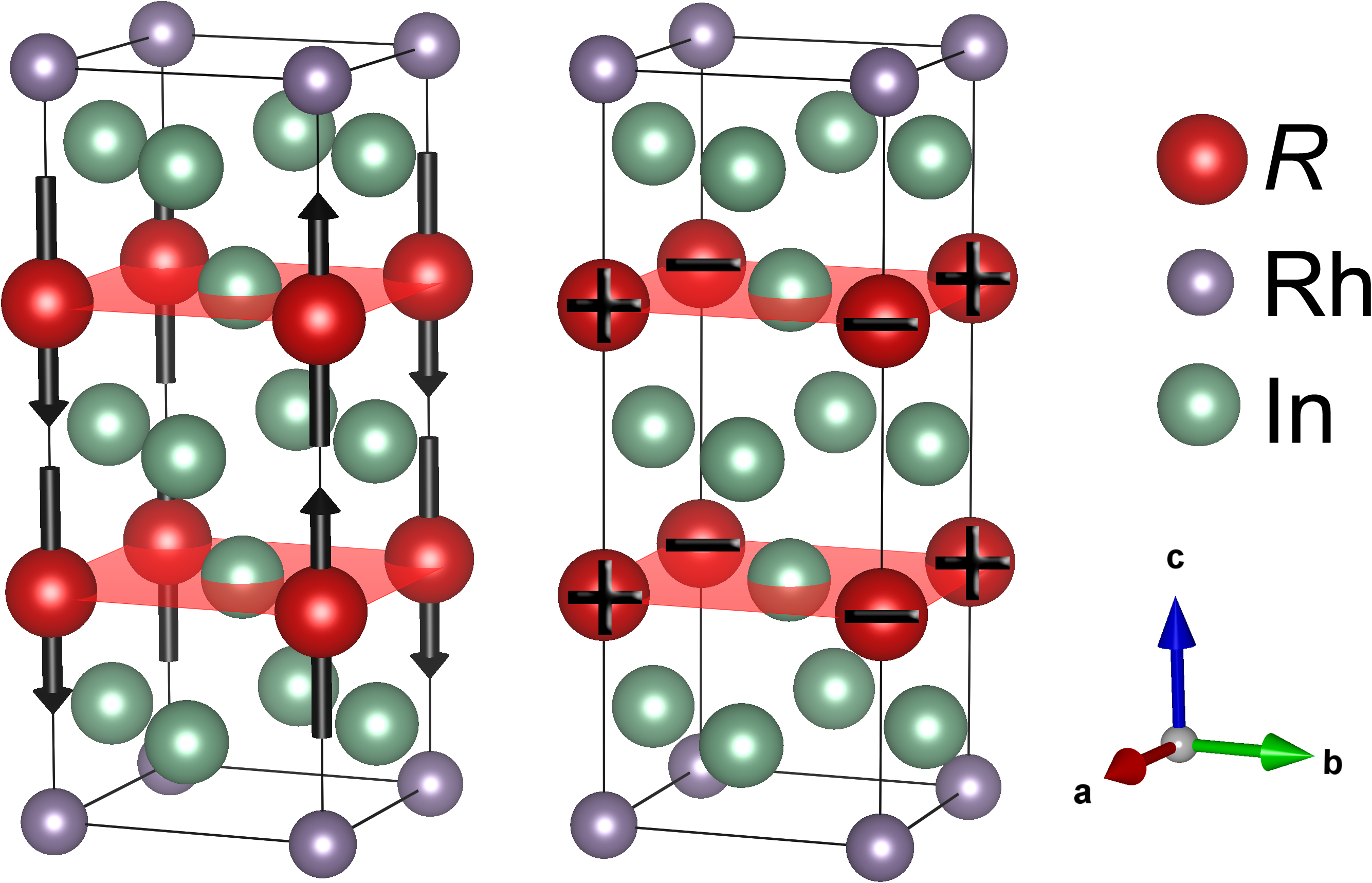}
}
\vspace{0mm} \caption{\label{magstr} Magnetic
structure of a) Nd$_2$RhIn$_8$ and Dy$_2$RhIn$_8$, b) Er$_2$RhIn$_8$ compounds.
Note that magnetic moments of Er$_2$RhIn$_8$ can point anywhere within the $ab$-plane,
but are all parallel to each other. }
\end{figure}

The obtained magnetic structures are depicted in
Fig.\ref{magstr}. The refined moments are summarized in Table \ref{parameters}.
The comparison of observed and calculated intensities for all compounds
is shown in Fig.\ref{Intensity}.
For completion of the magnetic refinement the rare-earth moments were allowed to lie in a general direction by combining two representations in order to check a lowering of symmetry.
We did not observe any noticeable improvement of the fits and the local minima
were always found within 1 - 2 degrees out of the previously determined direction using a single representation.
We can therefore conclude that the magnetic moments of Nd$_2$RhIn$_8$ and Dy$_2$RhIn$_8$
lie along the tetragonal $c$-axis, maintaining the tetragonal symmetry in the group \emph{P$_I$4/mm'm'}, while they lie within the $ab$-plane in the case of Er$_2$RhIn$_8$, lowering the symmetry at least to orthorhombic (remember that for experimental limitations we could not
determine the directions of the moment within the $ab$-plane).

The temperature dependence of the intensity of the (1/2, 1/2, 1/2) magnetic reflection
for each compound is shown in Fig. \ref{Tdep}. The data were fitted to the power law

\begin{equation}  \label{crit}
I \propto (T_N-T)^{2\beta}.
\end{equation}

The determined transition temperatures $T_N$ as well as the critical exponents $\beta$ are listed in Table \ref{parameters}.
These experimental results are incompatible with Ising prediction
for the three-dimensional $\beta \sim 0.313$ and for the two-dimensional $\beta = 0.125$ systems.
However, both neodymium and dysprosium compounds reveal qualitatively similar critical behavior pointing to an identical ordering mechanism.
Er$_2$RhIn$_8$ ordering coefficient suggests a more pronounced two-dimensional character.
The small value of $\beta = 0.16$ is rather different from the value $\beta = 0.33$ determined for isostructural Er$_2$CoGa$_8$ \cite{ErTmCoGa218}.
We observe a significantly steeper increase of the spontaneous magnetization compared to the gallium compound, despite both materials share a similar magnetic structure.

Let us now compare our results with the magnetic structures in related compounds in terms of dimensionality.
As mentioned in the introduction, the "218" compounds can be seen as transition from the nearly two-dimensional "115" towards the three-dimensional "13" compounds.
In the neodymium compounds the different "13" and "115" magnetic structures were
ascribed to competing (NdIn$_3$) or matching (NdRhIn$_5$)
crystal-field and exchange anisotropies \cite{NdRh115}.
The magnetic moments in both Nd$_2$RhIn$_8$ and NdRhIn$_5$ point
along the c-axis, driven by the crystal-field anisotropy.
The coupling between the neighboring Nd moments is antiferromagnetic
within the basal planes, although the moments propagate differently:
$\textbf{k}_{in-plane}$ = (1/2, 1/2) in Nd$_2$RhIn$_8$ and $\textbf{k}_{in-plane}$ =
(1/2, 0) in NdRhIn$_5$. The NdIn$_3$
layer (in NdRhIn$_5$) or bilayer (in Nd$_2$RhIn$_8$) is then
separated by a RhIn$_2$ layer. The Nd-Nd coupling along the c-axis
across this non-magnetic layer is in both cases also
antiferromagnetic. The coupling along the c-axis within the cubic
NdIn$_3$ blocks in Nd$_2$RhIn$_8$ is ferromagnetic, i.e. these
cubic blocks form the same magnetic structure occuring in the
ground state of NdIn$_3$. The magnetic structure can be viewed
also in the following way: among the two nearest Nd layers it acts
exactly as in NdIn$_3$ (diagonal propagation in the plane
perpendicular to the moments) while another Nd bilayer, separated
by RhIn$_2$ layer, is coupled antiferromagnetically creating the
overall propagation vector $\textbf{k}$ = (1/2, 1/2, 1/2).

Similar conclusions are valid for dysprosium compounds,
except the fact that in the cubic DyIn$_3$ the
magnetic moments point out of the main crystallographic directions.
Recently studied gallium analogue of the dysprosium compound, Dy$_2$CoGa$_8$, shows the same
magnetic structure and stacking along the $c$-axis \cite{GdTbDyCoGa218}.
Stacking of moments along the $c$-axis $++--$ in Nd$_2$RhIn$_8$ and Dy$_2$RhIn$_8$
is different from the stacking $+-+-$ revealed for Tb$_2$RhIn$_8$ \cite{TbRh218}.
This is then reflected in the qualitatively different magnetization
curves in magnetic fields above 10~T applied along the $a$-axis \cite{TbNd-high-field}.

No magnetic structure is reported for any of the erbium 115 compounds.
We can compare our results to the gallium analogue Er$_2$CoGa$_8$,
which has $\textbf{k}$ = (0, 1/2, 0), i.e. it propagates only along the
direction of the magnetic moments with $+-+-$ stacking along the $c$-axis.
This qualitative change of stacking within the unit cell
as well as different propagation vector is probably caused by the smaller distance
between Er atoms in the gallium compound (4.2287 $\textrm{\AA}$ in Er$_2$CoGa$_8$ \cite{ErTmCoGa218} compared to 4.5284 $\textrm{\AA}$ in Er$_2$RhIn$_8$).
The determined amplitude of the magnetic moment in the gallium compound 4.7~$\mu_B$ is also significantly reduced
in comparison with 6.4~$\mu_B$ for its indium relative.

In all three compounds the amplitude of the ordered moments is reduced from the expected values of the free ion,
in agreement with other compounds from the series \cite{ErTmCoGa218,RRh115-thesis}.
This is typical
for tetragonal CEF driven magnetic structures \cite{szytula1994handbook}, like for example
DyCo$_2$Si$_2$ \cite{DyCo2Si2-Iwata1990}.
Interpolating the measured magnetization curves along the $c$-axis
to zero magnetic field for Nd$_2$RhIn$_8$ and Dy$_2$RhIn$_8$ gives the values of
2.2 and 7.2 $\mu_B$ per $R$ atom, respectively \cite{TbNd-high-field,RRh218-our},
which are in good agreement with our experimental values.
Doing the same for Er$_2$RhIn$_8$ leads to the value of 7.8 $\mu_B$ per Er for the magnetic field applied along the [110]
direction and to the value of 6.9 $\mu_B$ per Er for the magnetic field applied along the [100] direction \cite{RRh218-our}.
From the determined Shubinikov groups is clear that \emph{irrep} $\Gamma_{10}$
is always connected with lowering of the symmetry and creation of the magnetic domains.
That is the reason, why values from bulk magnetization measurements are bigger than
the value of 6.4 $\mu_B$ obtained from the neutron diffraction.

The magnetic structures in the corresponding cerium compounds are
more complex.
An incommensurate spiral structure, with Ce moments
within the $ab$-planes, is observed in CeRhIn$_5$ \cite{CeRh115-MagStruct}. The
magnetic structure of Ce$_2$RhIn$_8$ is described by the propagation vector
$\textbf{k}$ = (1/2, 1/2, 0) and Ce moments pointing $38^{\circ}$ out of the
tetragonal $c$-axis. The coupling within the basal planes is the
same as in Nd$_2$RhIn$_8$, but the coupling along the $c$-axis is
different: it is antiferromagnetic across the non-magnetic RhIn$_2$
layer as well as within the cubic CeIn$_3$ blocks. The main
difference is however the moment direction.
Thus, the resulting structure lowers the symmetry by mixing two representations
within the same exchange multiplet \cite{Exchange_Multiplets}.
We assume that this is the consequence of stronger isotropic exchange interactions
with respect to the anisotropy in the Ce compound.

\section{Conclusion}

We have determined the magnetic structure of three intermetallic
compounds, Nd$_2$RhIn$_8$, Dy$_2$RhIn$_8$ and Er$_2$RhIn$_8$, by the means of
neutron diffraction experiments. All compounds are characterized by the propagation vector $\textbf{k}$ = (1/2, 1/2, 1/2) with ferromagnetic coupling between the nearest neighboring
rare-earth layers within the unit cell.
The magnetic moment direction reflects the crystal-field anisotropy in these compounds.
The magnetic moments of
Nd$_2$RhIn$_8$ and Dy$_2$RhIn$_8$ lie along the $c$-axis, while the moment of Er$_2$RhIn$_8$ lies within
the $ab$-plane, reaching values of 2.53, 6.9 and 6.4 $\mu_B$, respectively.

\begin{acknowledgments}
This work was supported by
the Czech Science Foundation under Grant No. P204-13-12227S and
by the Grant Agency of Charles University under the Grant No.
348511. The work is a part of research project LG14037 financed by the
Ministry of Education of Czech Republic. Samples were prepared in MLTL (http://mltl.eu/),
which is supported within the program of Czech Research Infrastructures (project No. LM2011025).
We acknowledge also the ILL for the beam-time allocation and support during our measurement.
\end{acknowledgments}

\bibliography{218-mag-struct}

\providecommand{\noopsort}[1]{}\providecommand{\singleletter}[1]{#1}%
\begin{thebibliography}{59}%
\makeatletter
\providecommand \@ifxundefined [1]{%
 \@ifx{#1\undefined}
}%
\providecommand \@ifnum [1]{%
 \ifnum #1\expandafter \@firstoftwo
 \else \expandafter \@secondoftwo
 \fi
}%
\providecommand \@ifx [1]{%
 \ifx #1\expandafter \@firstoftwo
 \else \expandafter \@secondoftwo
 \fi
}%
\providecommand \natexlab [1]{#1}%
\providecommand \enquote  [1]{``#1''}%
\providecommand \bibnamefont  [1]{#1}%
\providecommand \bibfnamefont [1]{#1}%
\providecommand \citenamefont [1]{#1}%
\providecommand \href@noop [0]{\@secondoftwo}%
\providecommand \href [0]{\begingroup \@sanitize@url \@href}%
\providecommand \@href[1]{\@@startlink{#1}\@@href}%
\providecommand \@@href[1]{\endgroup#1\@@endlink}%
\providecommand \@sanitize@url [0]{\catcode `\\12\catcode `\$12\catcode
  `\&12\catcode `\#12\catcode `\^12\catcode `\_12\catcode `\%12\relax}%
\providecommand \@@startlink[1]{}%
\providecommand \@@endlink[0]{}%
\providecommand \url  [0]{\begingroup\@sanitize@url \@url }%
\providecommand \@url [1]{\endgroup\@href {#1}{\urlprefix }}%
\providecommand \urlprefix  [0]{URL }%
\providecommand \Eprint [0]{\href }%
\providecommand \doibase [0]{http://dx.doi.org/}%
\providecommand \selectlanguage [0]{\@gobble}%
\providecommand \bibinfo  [0]{\@secondoftwo}%
\providecommand \bibfield  [0]{\@secondoftwo}%
\providecommand \translation [1]{[#1]}%
\providecommand \BibitemOpen [0]{}%
\providecommand \bibitemStop [0]{}%
\providecommand \bibitemNoStop [0]{.\EOS\space}%
\providecommand \EOS [0]{\spacefactor3000\relax}%
\providecommand \BibitemShut  [1]{\csname bibitem#1\endcsname}%
\let\auto@bib@innerbib\@empty
%</preamble>
\bibitem [{\citenamefont {Hegger}\ \emph {et~al.}(2000)\citenamefont {Hegger},
  \citenamefont {Petrovic}, \citenamefont {Moshopoulou}, \citenamefont
  {Hundley}, \citenamefont {Sarrao}, \citenamefont {Fisk},\ and\ \citenamefont
  {Thompson}}]{CeRh115-SC}%
  \BibitemOpen
  \bibfield  {author} {\bibinfo {author} {\bibfnamefont {H.}~\bibnamefont
  {Hegger}}, \bibinfo {author} {\bibfnamefont {C.}~\bibnamefont {Petrovic}},
  \bibinfo {author} {\bibfnamefont {E.~G.}\ \bibnamefont {Moshopoulou}},
  \bibinfo {author} {\bibfnamefont {M.~F.}\ \bibnamefont {Hundley}}, \bibinfo
  {author} {\bibfnamefont {J.~L.}\ \bibnamefont {Sarrao}}, \bibinfo {author}
  {\bibfnamefont {Z.}~\bibnamefont {Fisk}}, \ and\ \bibinfo {author}
  {\bibfnamefont {J.~D.}\ \bibnamefont {Thompson}},\ }\href {\doibase
  10.1103/PhysRevLett.84.4986} {\bibfield  {journal} {\bibinfo  {journal}
  {Phys. Rev. Lett.}\ }\textbf {\bibinfo {volume} {84}},\ \bibinfo {pages}
  {4986} (\bibinfo {year} {2000})}\BibitemShut {NoStop}%
\bibitem [{\citenamefont {Petrovic}\ \emph
  {et~al.}(2001{\natexlab{a}})\citenamefont {Petrovic}, \citenamefont
  {Pagliuso}, \citenamefont {Hundley}, \citenamefont {Movshovich},
  \citenamefont {Sarrao}, \citenamefont {Thompson}, \citenamefont {Fisk},\ and\
  \citenamefont {Monthoux}}]{CeCo115-SC}%
  \BibitemOpen
  \bibfield  {author} {\bibinfo {author} {\bibfnamefont {C.}~\bibnamefont
  {Petrovic}}, \bibinfo {author} {\bibfnamefont {P.~G.}\ \bibnamefont
  {Pagliuso}}, \bibinfo {author} {\bibfnamefont {M.~F.}\ \bibnamefont
  {Hundley}}, \bibinfo {author} {\bibfnamefont {R.}~\bibnamefont {Movshovich}},
  \bibinfo {author} {\bibfnamefont {J.~L.}\ \bibnamefont {Sarrao}}, \bibinfo
  {author} {\bibfnamefont {J.~D.}\ \bibnamefont {Thompson}}, \bibinfo {author}
  {\bibfnamefont {Z.}~\bibnamefont {Fisk}}, \ and\ \bibinfo {author}
  {\bibfnamefont {P.}~\bibnamefont {Monthoux}},\ }\href
  {http://stacks.iop.org/0953-8984/13/i=17/a=103} {\bibfield  {journal}
  {\bibinfo  {journal} {Journal of Physics: Condensed Matter}\ }\textbf
  {\bibinfo {volume} {13}},\ \bibinfo {pages} {L337} (\bibinfo {year}
  {2001}{\natexlab{a}})}\BibitemShut {NoStop}%
\bibitem [{\citenamefont {Petrovic}\ \emph
  {et~al.}(2001{\natexlab{b}})\citenamefont {Petrovic}, \citenamefont
  {Movshovich}, \citenamefont {Jaime}, \citenamefont {Pagliuso}, \citenamefont
  {Hundley}, \citenamefont {Sarrao}, \citenamefont {Fisk},\ and\ \citenamefont
  {Thompson}}]{CeIr115-SC}%
  \BibitemOpen
  \bibfield  {author} {\bibinfo {author} {\bibfnamefont {C.}~\bibnamefont
  {Petrovic}}, \bibinfo {author} {\bibfnamefont {R.}~\bibnamefont
  {Movshovich}}, \bibinfo {author} {\bibfnamefont {M.}~\bibnamefont {Jaime}},
  \bibinfo {author} {\bibfnamefont {P.~G.}\ \bibnamefont {Pagliuso}}, \bibinfo
  {author} {\bibfnamefont {M.~F.}\ \bibnamefont {Hundley}}, \bibinfo {author}
  {\bibfnamefont {J.~L.}\ \bibnamefont {Sarrao}}, \bibinfo {author}
  {\bibfnamefont {Z.}~\bibnamefont {Fisk}}, \ and\ \bibinfo {author}
  {\bibfnamefont {J.~D.}\ \bibnamefont {Thompson}},\ }\href
  {http://stacks.iop.org/0295-5075/53/i=3/a=354} {\bibfield  {journal}
  {\bibinfo  {journal} {EPL (Europhysics Letters)}\ }\textbf {\bibinfo {volume}
  {53}},\ \bibinfo {pages} {354} (\bibinfo {year}
  {2001}{\natexlab{b}})}\BibitemShut {NoStop}%
\bibitem [{\citenamefont {Kaczorowski}\ \emph {et~al.}(2010)\citenamefont
  {Kaczorowski}, \citenamefont {Gnida}, \citenamefont {Pikul},\ and\
  \citenamefont {Tran}}]{CePd218}%
  \BibitemOpen
  \bibfield  {author} {\bibinfo {author} {\bibfnamefont {D.}~\bibnamefont
  {Kaczorowski}}, \bibinfo {author} {\bibfnamefont {D.}~\bibnamefont {Gnida}},
  \bibinfo {author} {\bibfnamefont {A.}~\bibnamefont {Pikul}}, \ and\ \bibinfo
  {author} {\bibfnamefont {V.}~\bibnamefont {Tran}},\ }\href {\doibase
  http://dx.doi.org/10.1016/j.ssc.2009.12.007} {\bibfield  {journal} {\bibinfo
  {journal} {Solid State Communications}\ }\textbf {\bibinfo {volume} {150}},\
  \bibinfo {pages} {411 } (\bibinfo {year} {2010})}\BibitemShut {NoStop}%
\bibitem [{\citenamefont {Thompson}\ and\ \citenamefont
  {Fisk}(2012)}]{Ce115-Fisk}%
  \BibitemOpen
  \bibfield  {author} {\bibinfo {author} {\bibfnamefont {J.~D.}\ \bibnamefont
  {Thompson}}\ and\ \bibinfo {author} {\bibfnamefont {Z.}~\bibnamefont
  {Fisk}},\ }\href {\doibase 10.1143/JPSJ.81.011002} {\bibfield  {journal}
  {\bibinfo  {journal} {Journal of the Physical Society of Japan}\ }\textbf
  {\bibinfo {volume} {81}},\ \bibinfo {pages} {011002} (\bibinfo {year}
  {2012})}\BibitemShut {NoStop}%
\bibitem [{\citenamefont {Ott}\ \emph {et~al.}(1984)\citenamefont {Ott},
  \citenamefont {Rudigier}, \citenamefont {Rice}, \citenamefont {Ueda},
  \citenamefont {Fisk},\ and\ \citenamefont {Smith}}]{HeAnalogy-Ott1984}%
  \BibitemOpen
  \bibfield  {author} {\bibinfo {author} {\bibfnamefont {H.~R.}\ \bibnamefont
  {Ott}}, \bibinfo {author} {\bibfnamefont {H.}~\bibnamefont {Rudigier}},
  \bibinfo {author} {\bibfnamefont {T.~M.}\ \bibnamefont {Rice}}, \bibinfo
  {author} {\bibfnamefont {K.}~\bibnamefont {Ueda}}, \bibinfo {author}
  {\bibfnamefont {Z.}~\bibnamefont {Fisk}}, \ and\ \bibinfo {author}
  {\bibfnamefont {J.~L.}\ \bibnamefont {Smith}},\ }\href {\doibase
  10.1103/PhysRevLett.52.1915} {\bibfield  {journal} {\bibinfo  {journal}
  {Phys. Rev. Lett.}\ }\textbf {\bibinfo {volume} {52}},\ \bibinfo {pages}
  {1915} (\bibinfo {year} {1984})}\BibitemShut {NoStop}%
\bibitem [{\citenamefont {Benoit}\ \emph {et~al.}(1980)\citenamefont {Benoit},
  \citenamefont {Boucherle}, \citenamefont {Convert}, \citenamefont {Flouquet},
  \citenamefont {Palleau},\ and\ \citenamefont {Schweizer}}]{Ce13}%
  \BibitemOpen
  \bibfield  {author} {\bibinfo {author} {\bibfnamefont {A.}~\bibnamefont
  {Benoit}}, \bibinfo {author} {\bibfnamefont {J.}~\bibnamefont {Boucherle}},
  \bibinfo {author} {\bibfnamefont {P.}~\bibnamefont {Convert}}, \bibinfo
  {author} {\bibfnamefont {J.}~\bibnamefont {Flouquet}}, \bibinfo {author}
  {\bibfnamefont {J.}~\bibnamefont {Palleau}}, \ and\ \bibinfo {author}
  {\bibfnamefont {J.}~\bibnamefont {Schweizer}},\ }\href {\doibase
  10.1016/0038-1098(80)90560-8} {\bibfield  {journal} {\bibinfo  {journal}
  {Solid State Communications}\ }\textbf {\bibinfo {volume} {34}},\ \bibinfo
  {pages} {293 } (\bibinfo {year} {1980})}\BibitemShut {NoStop}%
\bibitem [{\citenamefont {Schenck}\ \emph {et~al.}(2004)\citenamefont
  {Schenck}, \citenamefont {Gygax}, \citenamefont {Ueda},\ and\ \citenamefont
  {Onuki}}]{CeRh218-MagStruct}%
  \BibitemOpen
  \bibfield  {author} {\bibinfo {author} {\bibfnamefont {A.}~\bibnamefont
  {Schenck}}, \bibinfo {author} {\bibfnamefont {F.~N.}\ \bibnamefont {Gygax}},
  \bibinfo {author} {\bibfnamefont {T.}~\bibnamefont {Ueda}}, \ and\ \bibinfo
  {author} {\bibfnamefont {Y.}~\bibnamefont {Onuki}},\ }\href {\doibase
  10.1103/PhysRevB.70.054415} {\bibfield  {journal} {\bibinfo  {journal} {Phys.
  Rev. B}\ }\textbf {\bibinfo {volume} {70}},\ \bibinfo {pages} {054415}
  (\bibinfo {year} {2004})}\BibitemShut {NoStop}%
\bibitem [{\citenamefont {Bao}\ \emph {et~al.}(2000)\citenamefont {Bao},
  \citenamefont {Pagliuso}, \citenamefont {Sarrao}, \citenamefont {Thompson},
  \citenamefont {Fisk}, \citenamefont {Lynn},\ and\ \citenamefont
  {Erwin}}]{CeRh115-MagStruct}%
  \BibitemOpen
  \bibfield  {author} {\bibinfo {author} {\bibfnamefont {W.}~\bibnamefont
  {Bao}}, \bibinfo {author} {\bibfnamefont {P.~G.}\ \bibnamefont {Pagliuso}},
  \bibinfo {author} {\bibfnamefont {J.~L.}\ \bibnamefont {Sarrao}}, \bibinfo
  {author} {\bibfnamefont {J.~D.}\ \bibnamefont {Thompson}}, \bibinfo {author}
  {\bibfnamefont {Z.}~\bibnamefont {Fisk}}, \bibinfo {author} {\bibfnamefont
  {J.~W.}\ \bibnamefont {Lynn}}, \ and\ \bibinfo {author} {\bibfnamefont
  {R.~W.}\ \bibnamefont {Erwin}},\ }\href {\doibase 10.1103/PhysRevB.62.R14621}
  {\bibfield  {journal} {\bibinfo  {journal} {Phys. Rev. B}\ }\textbf {\bibinfo
  {volume} {62}},\ \bibinfo {pages} {R14621} (\bibinfo {year}
  {2000})}\BibitemShut {NoStop}%
\bibitem [{\citenamefont {Bao}\ \emph {et~al.}(2003)\citenamefont {Bao},
  \citenamefont {Pagliuso}, \citenamefont {Sarrao}, \citenamefont {Thompson},
  \citenamefont {Fisk}, \citenamefont {Lynn},\ and\ \citenamefont
  {Erwin}}]{CeRh115-MagStruct-errata}%
  \BibitemOpen
  \bibfield  {author} {\bibinfo {author} {\bibfnamefont {W.}~\bibnamefont
  {Bao}}, \bibinfo {author} {\bibfnamefont {P.~G.}\ \bibnamefont {Pagliuso}},
  \bibinfo {author} {\bibfnamefont {J.~L.}\ \bibnamefont {Sarrao}}, \bibinfo
  {author} {\bibfnamefont {J.~D.}\ \bibnamefont {Thompson}}, \bibinfo {author}
  {\bibfnamefont {Z.}~\bibnamefont {Fisk}}, \bibinfo {author} {\bibfnamefont
  {J.~W.}\ \bibnamefont {Lynn}}, \ and\ \bibinfo {author} {\bibfnamefont
  {R.~W.}\ \bibnamefont {Erwin}},\ }\href {\doibase 10.1103/PhysRevB.67.099903}
  {\bibfield  {journal} {\bibinfo  {journal} {Phys. Rev. B}\ }\textbf {\bibinfo
  {volume} {67}},\ \bibinfo {pages} {099903(E)} (\bibinfo {year}
  {2003})}\BibitemShut {NoStop}%
\bibitem [{\citenamefont {Christianson}\ \emph {et~al.}(2002)\citenamefont
  {Christianson}, \citenamefont {Lawrence}, \citenamefont {Pagliuso},
  \citenamefont {Moreno}, \citenamefont {Sarrao}, \citenamefont {Thompson},
  \citenamefont {Riseborough}, \citenamefont {Kern}, \citenamefont
  {Goremychkin},\ and\ \citenamefont {Lacerda}}]{CeRh115-CF}%
  \BibitemOpen
  \bibfield  {author} {\bibinfo {author} {\bibfnamefont {A.~D.}\ \bibnamefont
  {Christianson}}, \bibinfo {author} {\bibfnamefont {J.~M.}\ \bibnamefont
  {Lawrence}}, \bibinfo {author} {\bibfnamefont {P.~G.}\ \bibnamefont
  {Pagliuso}}, \bibinfo {author} {\bibfnamefont {N.~O.}\ \bibnamefont
  {Moreno}}, \bibinfo {author} {\bibfnamefont {J.~L.}\ \bibnamefont {Sarrao}},
  \bibinfo {author} {\bibfnamefont {J.~D.}\ \bibnamefont {Thompson}}, \bibinfo
  {author} {\bibfnamefont {P.~S.}\ \bibnamefont {Riseborough}}, \bibinfo
  {author} {\bibfnamefont {S.}~\bibnamefont {Kern}}, \bibinfo {author}
  {\bibfnamefont {E.~A.}\ \bibnamefont {Goremychkin}}, \ and\ \bibinfo {author}
  {\bibfnamefont {A.~H.}\ \bibnamefont {Lacerda}},\ }\href {\doibase
  10.1103/PhysRevB.66.193102} {\bibfield  {journal} {\bibinfo  {journal} {Phys.
  Rev. B}\ }\textbf {\bibinfo {volume} {66}},\ \bibinfo {pages} {193102}
  (\bibinfo {year} {2002})}\BibitemShut {NoStop}%
\bibitem [{\citenamefont {Kenzelmann}\ \emph {et~al.}(2008)\citenamefont
  {Kenzelmann}, \citenamefont {Str\"{a}ssle}, \citenamefont {Niedermayer},
  \citenamefont {Sigrist}, \citenamefont {Padmanabhan}, \citenamefont
  {Zolliker}, \citenamefont {Bianchi}, \citenamefont {Movshovich},
  \citenamefont {Bauer}, \citenamefont {Sarrao},\ and\ \citenamefont
  {Thompson}}]{CeCo115-MagStruct}%
  \BibitemOpen
  \bibfield  {author} {\bibinfo {author} {\bibfnamefont {M.}~\bibnamefont
  {Kenzelmann}}, \bibinfo {author} {\bibfnamefont {T.}~\bibnamefont
  {Str\"{a}ssle}}, \bibinfo {author} {\bibfnamefont {C.}~\bibnamefont
  {Niedermayer}}, \bibinfo {author} {\bibfnamefont {M.}~\bibnamefont
  {Sigrist}}, \bibinfo {author} {\bibfnamefont {B.}~\bibnamefont
  {Padmanabhan}}, \bibinfo {author} {\bibfnamefont {M.}~\bibnamefont
  {Zolliker}}, \bibinfo {author} {\bibfnamefont {A.~D.}\ \bibnamefont
  {Bianchi}}, \bibinfo {author} {\bibfnamefont {R.}~\bibnamefont {Movshovich}},
  \bibinfo {author} {\bibfnamefont {E.~D.}\ \bibnamefont {Bauer}}, \bibinfo
  {author} {\bibfnamefont {J.~L.}\ \bibnamefont {Sarrao}}, \ and\ \bibinfo
  {author} {\bibfnamefont {J.~D.}\ \bibnamefont {Thompson}},\ }\href {\doibase
  10.1126/science.1161818} {\bibfield  {journal} {\bibinfo  {journal}
  {Science}\ }\textbf {\bibinfo {volume} {321}},\ \bibinfo {pages} {1652}
  (\bibinfo {year} {2008})}\BibitemShut {NoStop}%
\bibitem [{\citenamefont {Raymond}\ \emph {et~al.}(2014)\citenamefont
  {Raymond}, \citenamefont {Ramos}, \citenamefont {Aoki}, \citenamefont
  {Knebel}, \citenamefont {Mineev},\ and\ \citenamefont
  {Lapertot}}]{CeNdCo-115-magstruct}%
  \BibitemOpen
  \bibfield  {author} {\bibinfo {author} {\bibfnamefont {S.}~\bibnamefont
  {Raymond}}, \bibinfo {author} {\bibfnamefont {S.~M.}\ \bibnamefont {Ramos}},
  \bibinfo {author} {\bibfnamefont {D.}~\bibnamefont {Aoki}}, \bibinfo {author}
  {\bibfnamefont {G.}~\bibnamefont {Knebel}}, \bibinfo {author} {\bibfnamefont
  {V.~P.}\ \bibnamefont {Mineev}}, \ and\ \bibinfo {author} {\bibfnamefont
  {G.}~\bibnamefont {Lapertot}},\ }\href {\doibase 10.7566/JPSJ.83.013707}
  {\bibfield  {journal} {\bibinfo  {journal} {Journal of the Physical Society
  of Japan}\ }\textbf {\bibinfo {volume} {83}},\ \bibinfo {pages} {013707}
  (\bibinfo {year} {2014})}\BibitemShut {NoStop}%
\bibitem [{\citenamefont {Bauer}\ \emph {et~al.}(2010)\citenamefont {Bauer},
  \citenamefont {Lee}, \citenamefont {Sidorov}, \citenamefont {Kurita},
  \citenamefont {Gofryk}, \citenamefont {Zhu}, \citenamefont {Ronning},
  \citenamefont {Movshovich}, \citenamefont {Thompson},\ and\ \citenamefont
  {Park}}]{CePt127}%
  \BibitemOpen
  \bibfield  {author} {\bibinfo {author} {\bibfnamefont {E.~D.}\ \bibnamefont
  {Bauer}}, \bibinfo {author} {\bibfnamefont {H.~O.}\ \bibnamefont {Lee}},
  \bibinfo {author} {\bibfnamefont {V.~A.}\ \bibnamefont {Sidorov}}, \bibinfo
  {author} {\bibfnamefont {N.}~\bibnamefont {Kurita}}, \bibinfo {author}
  {\bibfnamefont {K.}~\bibnamefont {Gofryk}}, \bibinfo {author} {\bibfnamefont
  {J.-X.}\ \bibnamefont {Zhu}}, \bibinfo {author} {\bibfnamefont
  {F.}~\bibnamefont {Ronning}}, \bibinfo {author} {\bibfnamefont
  {R.}~\bibnamefont {Movshovich}}, \bibinfo {author} {\bibfnamefont {J.~D.}\
  \bibnamefont {Thompson}}, \ and\ \bibinfo {author} {\bibfnamefont
  {T.}~\bibnamefont {Park}},\ }\href {\doibase 10.1103/PhysRevB.81.180507}
  {\bibfield  {journal} {\bibinfo  {journal} {Phys. Rev. B}\ }\textbf {\bibinfo
  {volume} {81}},\ \bibinfo {pages} {180507} (\bibinfo {year}
  {2010})}\BibitemShut {NoStop}%
\bibitem [{\citenamefont {Sakai}\ \emph {et~al.}(2011)\citenamefont {Sakai},
  \citenamefont {Tokunaga}, \citenamefont {Kambe}, \citenamefont {Lee},
  \citenamefont {Sidorov}, \citenamefont {Tobash}, \citenamefont {Ronning},
  \citenamefont {Bauer},\ and\ \citenamefont {Thompson}}]{CePt127-NMR}%
  \BibitemOpen
  \bibfield  {author} {\bibinfo {author} {\bibfnamefont {H.}~\bibnamefont
  {Sakai}}, \bibinfo {author} {\bibfnamefont {Y.}~\bibnamefont {Tokunaga}},
  \bibinfo {author} {\bibfnamefont {S.}~\bibnamefont {Kambe}}, \bibinfo
  {author} {\bibfnamefont {H.-O.}\ \bibnamefont {Lee}}, \bibinfo {author}
  {\bibfnamefont {V.~A.}\ \bibnamefont {Sidorov}}, \bibinfo {author}
  {\bibfnamefont {P.~H.}\ \bibnamefont {Tobash}}, \bibinfo {author}
  {\bibfnamefont {F.}~\bibnamefont {Ronning}}, \bibinfo {author} {\bibfnamefont
  {E.~D.}\ \bibnamefont {Bauer}}, \ and\ \bibinfo {author} {\bibfnamefont
  {J.~D.}\ \bibnamefont {Thompson}},\ }\href {\doibase
  10.1103/PhysRevB.83.140408} {\bibfield  {journal} {\bibinfo  {journal} {Phys.
  Rev. B}\ }\textbf {\bibinfo {volume} {83}},\ \bibinfo {pages} {140408}
  (\bibinfo {year} {2011})}\BibitemShut {NoStop}%
\bibitem [{\citenamefont {Amara}\ \emph {et~al.}(1994)\citenamefont {Amara},
  \citenamefont {Gal'{e}ra}, \citenamefont {Morin}, \citenamefont {Veres},\
  and\ \citenamefont {Burlet}}]{Nd13}%
  \BibitemOpen
  \bibfield  {author} {\bibinfo {author} {\bibfnamefont {M.}~\bibnamefont
  {Amara}}, \bibinfo {author} {\bibfnamefont {R.}~\bibnamefont {Gal'{e}ra}},
  \bibinfo {author} {\bibfnamefont {P.}~\bibnamefont {Morin}}, \bibinfo
  {author} {\bibfnamefont {T.}~\bibnamefont {Veres}}, \ and\ \bibinfo {author}
  {\bibfnamefont {P.}~\bibnamefont {Burlet}},\ }\href {\doibase
  10.1016/0304-8853(94)90665-3} {\bibfield  {journal} {\bibinfo  {journal}
  {Journal of Magnetism and Magnetic Materials}\ }\textbf {\bibinfo {volume}
  {130}},\ \bibinfo {pages} {127 } (\bibinfo {year} {1994})}\BibitemShut
  {NoStop}%
\bibitem [{\citenamefont {Kaczorowski}\ \emph {et~al.}(2011)\citenamefont
  {Kaczorowski}, \citenamefont {Belan}, \citenamefont {Sojka},\ and\
  \citenamefont {Kalychak}}]{PrPd218}%
  \BibitemOpen
  \bibfield  {author} {\bibinfo {author} {\bibfnamefont {D.}~\bibnamefont
  {Kaczorowski}}, \bibinfo {author} {\bibfnamefont {B.}~\bibnamefont {Belan}},
  \bibinfo {author} {\bibfnamefont {L.}~\bibnamefont {Sojka}}, \ and\ \bibinfo
  {author} {\bibfnamefont {Y.}~\bibnamefont {Kalychak}},\ }\href {\doibase
  http://dx.doi.org/10.1016/j.jallcom.2010.11.207} {\bibfield  {journal}
  {\bibinfo  {journal} {Journal of Alloys and Compounds}\ }\textbf {\bibinfo
  {volume} {509}},\ \bibinfo {pages} {3208 } (\bibinfo {year}
  {2011})}\BibitemShut {NoStop}%
\bibitem [{\citenamefont {Johnson}\ \emph {et~al.}(2010)\citenamefont
  {Johnson}, \citenamefont {Frawley}, \citenamefont {Manuel}, \citenamefont
  {Khalyavin}, \citenamefont {Adriano}, \citenamefont {Giles}, \citenamefont
  {Pagliuso},\ and\ \citenamefont {Hatton}}]{ErTmCoGa218}%
  \BibitemOpen
  \bibfield  {author} {\bibinfo {author} {\bibfnamefont {R.~D.}\ \bibnamefont
  {Johnson}}, \bibinfo {author} {\bibfnamefont {T.}~\bibnamefont {Frawley}},
  \bibinfo {author} {\bibfnamefont {P.}~\bibnamefont {Manuel}}, \bibinfo
  {author} {\bibfnamefont {D.~D.}\ \bibnamefont {Khalyavin}}, \bibinfo {author}
  {\bibfnamefont {C.}~\bibnamefont {Adriano}}, \bibinfo {author} {\bibfnamefont
  {C.}~\bibnamefont {Giles}}, \bibinfo {author} {\bibfnamefont {P.~G.}\
  \bibnamefont {Pagliuso}}, \ and\ \bibinfo {author} {\bibfnamefont {P.~D.}\
  \bibnamefont {Hatton}},\ }\href {\doibase 10.1103/PhysRevB.82.104407}
  {\bibfield  {journal} {\bibinfo  {journal} {Phys. Rev. B}\ }\textbf {\bibinfo
  {volume} {82}},\ \bibinfo {pages} {104407} (\bibinfo {year}
  {2010})}\BibitemShut {NoStop}%
\bibitem [{\citenamefont {Adriano}\ \emph {et~al.}(2009)\citenamefont
  {Adriano}, \citenamefont {Giles}, \citenamefont {Coelho}, \citenamefont
  {Faria},\ and\ \citenamefont {Pagliuso}}]{HoCoGa218}%
  \BibitemOpen
  \bibfield  {author} {\bibinfo {author} {\bibfnamefont {C.}~\bibnamefont
  {Adriano}}, \bibinfo {author} {\bibfnamefont {C.}~\bibnamefont {Giles}},
  \bibinfo {author} {\bibfnamefont {L.}~\bibnamefont {Coelho}}, \bibinfo
  {author} {\bibfnamefont {G.}~\bibnamefont {Faria}}, \ and\ \bibinfo {author}
  {\bibfnamefont {P.}~\bibnamefont {Pagliuso}},\ }\href {\doibase
  10.1016/j.physb.2009.07.127} {\bibfield  {journal} {\bibinfo  {journal}
  {Physica B: Condensed Matter}\ }\textbf {\bibinfo {volume} {404}},\ \bibinfo
  {pages} {3289 } (\bibinfo {year} {2009})},\ \bibinfo {note} {proceedings of
  the International Conference on Strongly Correlated Electron
  Systems}\BibitemShut {NoStop}%
\bibitem [{\citenamefont {Mardegan}\ \emph {et~al.}(2014)\citenamefont
  {Mardegan}, \citenamefont {Adriano}, \citenamefont {Vescovi}, \citenamefont
  {Faria}, \citenamefont {Pagliuso},\ and\ \citenamefont
  {Giles}}]{GdTbDyCoGa218}%
  \BibitemOpen
  \bibfield  {author} {\bibinfo {author} {\bibfnamefont {J.~R.~L.}\
  \bibnamefont {Mardegan}}, \bibinfo {author} {\bibfnamefont {C.}~\bibnamefont
  {Adriano}}, \bibinfo {author} {\bibfnamefont {R.~F.~C.}\ \bibnamefont
  {Vescovi}}, \bibinfo {author} {\bibfnamefont {G.~A.}\ \bibnamefont {Faria}},
  \bibinfo {author} {\bibfnamefont {P.~G.}\ \bibnamefont {Pagliuso}}, \ and\
  \bibinfo {author} {\bibfnamefont {C.}~\bibnamefont {Giles}},\ }\href
  {\doibase 10.1103/PhysRevB.89.115103} {\bibfield  {journal} {\bibinfo
  {journal} {Phys. Rev. B}\ }\textbf {\bibinfo {volume} {89}},\ \bibinfo
  {pages} {115103} (\bibinfo {year} {2014})}\BibitemShut {NoStop}%
\bibitem [{\citenamefont {Malachias}\ \emph {et~al.}(2008)\citenamefont
  {Malachias}, \citenamefont {Granado}, \citenamefont {Lora-Serrano},
  \citenamefont {Pagliuso},\ and\ \citenamefont {P\'erez}}]{Gd13}%
  \BibitemOpen
  \bibfield  {author} {\bibinfo {author} {\bibfnamefont {A.}~\bibnamefont
  {Malachias}}, \bibinfo {author} {\bibfnamefont {E.}~\bibnamefont {Granado}},
  \bibinfo {author} {\bibfnamefont {R.}~\bibnamefont {Lora-Serrano}}, \bibinfo
  {author} {\bibfnamefont {P.~G.}\ \bibnamefont {Pagliuso}}, \ and\ \bibinfo
  {author} {\bibfnamefont {C.~A.}\ \bibnamefont {P\'erez}},\ }\href {\doibase
  10.1103/PhysRevB.77.094425} {\bibfield  {journal} {\bibinfo  {journal} {Phys.
  Rev. B}\ }\textbf {\bibinfo {volume} {77}},\ \bibinfo {pages} {094425}
  (\bibinfo {year} {2008})}\BibitemShut {NoStop}%
\bibitem [{\citenamefont {Nereson}\ and\ \citenamefont
  {Arnold}(1970)}]{Tb13-Ho13}%
  \BibitemOpen
  \bibfield  {author} {\bibinfo {author} {\bibfnamefont {N.}~\bibnamefont
  {Nereson}}\ and\ \bibinfo {author} {\bibfnamefont {G.}~\bibnamefont
  {Arnold}},\ }\href {\doibase 10.1063/1.1674407} {\bibfield  {journal}
  {\bibinfo  {journal} {The Journal of Chemical Physics}\ }\textbf {\bibinfo
  {volume} {53}},\ \bibinfo {pages} {2818} (\bibinfo {year}
  {1970})}\BibitemShut {NoStop}%
\bibitem [{\citenamefont {Arnold}\ and\ \citenamefont {Nereson}(1969)}]{Dy13}%
  \BibitemOpen
  \bibfield  {author} {\bibinfo {author} {\bibfnamefont {G.}~\bibnamefont
  {Arnold}}\ and\ \bibinfo {author} {\bibfnamefont {N.}~\bibnamefont
  {Nereson}},\ }\href {\doibase 10.1063/1.1672201} {\bibfield  {journal}
  {\bibinfo  {journal} {The Journal of Chemical Physics}\ }\textbf {\bibinfo
  {volume} {51}},\ \bibinfo {pages} {1495} (\bibinfo {year}
  {1969})}\BibitemShut {NoStop}%
\bibitem [{\citenamefont {Czopnik}\ \emph {et~al.}(1988)\citenamefont
  {Czopnik}, \citenamefont {Madge},\ and\ \citenamefont {Stalinski}}]{Er13}%
  \BibitemOpen
  \bibfield  {author} {\bibinfo {author} {\bibfnamefont {A.}~\bibnamefont
  {Czopnik}}, \bibinfo {author} {\bibfnamefont {H.}~\bibnamefont {Madge}}, \
  and\ \bibinfo {author} {\bibfnamefont {B.}~\bibnamefont {Stalinski}},\ }\href
  {\doibase 10.1002/pssa.2211070255} {\bibfield  {journal} {\bibinfo  {journal}
  {physica status solidi (a)}\ }\textbf {\bibinfo {volume} {107}},\ \bibinfo
  {pages} {K151} (\bibinfo {year} {1988})}\BibitemShut {NoStop}%
\bibitem [{\citenamefont {Murasik}\ \emph {et~al.}(2002)\citenamefont
  {Murasik}, \citenamefont {Czopnik}, \citenamefont {Keller},\ and\
  \citenamefont {Konter}}]{Tm13}%
  \BibitemOpen
  \bibfield  {author} {\bibinfo {author} {\bibfnamefont {A.}~\bibnamefont
  {Murasik}}, \bibinfo {author} {\bibfnamefont {A.}~\bibnamefont {Czopnik}},
  \bibinfo {author} {\bibfnamefont {L.}~\bibnamefont {Keller}}, \ and\ \bibinfo
  {author} {\bibfnamefont {T.}~\bibnamefont {Konter}},\ }\href {\doibase
  10.1002/1521-396X(200202)189:2<R7::AID-PSSA99997>3.0.CO;2-3} {\bibfield
  {journal} {\bibinfo  {journal} {physica status solidi (a)}\ }\textbf
  {\bibinfo {volume} {189}},\ \bibinfo {pages} {R7} (\bibinfo {year}
  {2002})}\BibitemShut {NoStop}%
\bibitem [{\citenamefont {Chang}\ \emph {et~al.}(2002)\citenamefont {Chang},
  \citenamefont {Pagliuso}, \citenamefont {Bao}, \citenamefont {Gardner},
  \citenamefont {Swainson}, \citenamefont {Sarrao},\ and\ \citenamefont
  {Nakotte}}]{NdRh115}%
  \BibitemOpen
  \bibfield  {author} {\bibinfo {author} {\bibfnamefont {S.}~\bibnamefont
  {Chang}}, \bibinfo {author} {\bibfnamefont {P.~G.}\ \bibnamefont {Pagliuso}},
  \bibinfo {author} {\bibfnamefont {W.}~\bibnamefont {Bao}}, \bibinfo {author}
  {\bibfnamefont {J.~S.}\ \bibnamefont {Gardner}}, \bibinfo {author}
  {\bibfnamefont {I.~P.}\ \bibnamefont {Swainson}}, \bibinfo {author}
  {\bibfnamefont {J.~L.}\ \bibnamefont {Sarrao}}, \ and\ \bibinfo {author}
  {\bibfnamefont {H.}~\bibnamefont {Nakotte}},\ }\href {\doibase
  10.1103/PhysRevB.66.132417} {\bibfield  {journal} {\bibinfo  {journal} {Phys.
  Rev. B}\ }\textbf {\bibinfo {volume} {66}},\ \bibinfo {pages} {132417}
  (\bibinfo {year} {2002})}\BibitemShut {NoStop}%
\bibitem [{\citenamefont {Granado}\ \emph {et~al.}(2006)\citenamefont
  {Granado}, \citenamefont {Uchoa}, \citenamefont {Malachias}, \citenamefont
  {Lora-Serrano}, \citenamefont {Pagliuso},\ and\ \citenamefont
  {Westfahl}}]{GdRh115}%
  \BibitemOpen
  \bibfield  {author} {\bibinfo {author} {\bibfnamefont {E.}~\bibnamefont
  {Granado}}, \bibinfo {author} {\bibfnamefont {B.}~\bibnamefont {Uchoa}},
  \bibinfo {author} {\bibfnamefont {A.}~\bibnamefont {Malachias}}, \bibinfo
  {author} {\bibfnamefont {R.}~\bibnamefont {Lora-Serrano}}, \bibinfo {author}
  {\bibfnamefont {P.~G.}\ \bibnamefont {Pagliuso}}, \ and\ \bibinfo {author}
  {\bibfnamefont {H.}~\bibnamefont {Westfahl}},\ }\href {\doibase
  10.1103/PhysRevB.74.214428} {\bibfield  {journal} {\bibinfo  {journal} {Phys.
  Rev. B}\ }\textbf {\bibinfo {volume} {74}},\ \bibinfo {pages} {214428}
  (\bibinfo {year} {2006})}\BibitemShut {NoStop}%
\bibitem [{\citenamefont {Lora-Serrano}\ \emph
  {et~al.}(2006{\natexlab{a}})\citenamefont {Lora-Serrano}, \citenamefont
  {Giles}, \citenamefont {Granado}, \citenamefont {Garcia}, \citenamefont
  {Miranda}, \citenamefont {Ag\"uero}, \citenamefont {Mendonca~Ferreira},
  \citenamefont {Duque},\ and\ \citenamefont {Pagliuso}}]{TbRh115}%
  \BibitemOpen
  \bibfield  {author} {\bibinfo {author} {\bibfnamefont {R.}~\bibnamefont
  {Lora-Serrano}}, \bibinfo {author} {\bibfnamefont {C.}~\bibnamefont {Giles}},
  \bibinfo {author} {\bibfnamefont {E.}~\bibnamefont {Granado}}, \bibinfo
  {author} {\bibfnamefont {D.~J.}\ \bibnamefont {Garcia}}, \bibinfo {author}
  {\bibfnamefont {E.}~\bibnamefont {Miranda}}, \bibinfo {author} {\bibfnamefont
  {O.}~\bibnamefont {Ag\"uero}}, \bibinfo {author} {\bibfnamefont
  {L.}~\bibnamefont {Mendonca~Ferreira}}, \bibinfo {author} {\bibfnamefont
  {J.~G.~S.}\ \bibnamefont {Duque}}, \ and\ \bibinfo {author} {\bibfnamefont
  {P.~G.}\ \bibnamefont {Pagliuso}},\ }\href {\doibase
  10.1103/PhysRevB.74.214404} {\bibfield  {journal} {\bibinfo  {journal} {Phys.
  Rev. B}\ }\textbf {\bibinfo {volume} {74}},\ \bibinfo {pages} {214404}
  (\bibinfo {year} {2006}{\natexlab{a}})}\BibitemShut {NoStop}%
\bibitem [{\citenamefont {Hieu}(2007)}]{RRh115-thesis}%
  \BibitemOpen
  \bibfield  {author} {\bibinfo {author} {\bibfnamefont {N.~V.}\ \bibnamefont
  {Hieu}},\ }\emph {\bibinfo {title} {Single Crystal Growth and Magnetic
  Properties of RRhIn$_5$ Compounds ( R: Rare Earths )}},\ \href@noop {} {Ph.D.
  thesis},\ \bibinfo  {school} {Department of Physics, Graduate School of
  Science, Osaka University, Japan} (\bibinfo {year} {2007})\BibitemShut
  {NoStop}%
\bibitem [{\citenamefont {Tokunaga}\ \emph {et~al.}(2011)\citenamefont
  {Tokunaga}, \citenamefont {Saito}, \citenamefont {Sakai}, \citenamefont
  {Kambe}, \citenamefont {Sanada}, \citenamefont {Watanuki}, \citenamefont
  {Suzuki}, \citenamefont {Kawasaki},\ and\ \citenamefont
  {Kishimoto}}]{TbCoGa115}%
  \BibitemOpen
  \bibfield  {author} {\bibinfo {author} {\bibfnamefont {Y.}~\bibnamefont
  {Tokunaga}}, \bibinfo {author} {\bibfnamefont {Y.}~\bibnamefont {Saito}},
  \bibinfo {author} {\bibfnamefont {H.}~\bibnamefont {Sakai}}, \bibinfo
  {author} {\bibfnamefont {S.}~\bibnamefont {Kambe}}, \bibinfo {author}
  {\bibfnamefont {N.}~\bibnamefont {Sanada}}, \bibinfo {author} {\bibfnamefont
  {R.}~\bibnamefont {Watanuki}}, \bibinfo {author} {\bibfnamefont
  {K.}~\bibnamefont {Suzuki}}, \bibinfo {author} {\bibfnamefont
  {Y.}~\bibnamefont {Kawasaki}}, \ and\ \bibinfo {author} {\bibfnamefont
  {Y.}~\bibnamefont {Kishimoto}},\ }\href {\doibase 10.1103/PhysRevB.84.214403}
  {\bibfield  {journal} {\bibinfo  {journal} {Phys. Rev. B}\ }\textbf {\bibinfo
  {volume} {84}},\ \bibinfo {pages} {214403} (\bibinfo {year}
  {2011})}\BibitemShut {NoStop}%
\bibitem [{\citenamefont {Giles}\ \emph {et~al.}(2011)\citenamefont {Giles},
  \citenamefont {Adriano}, \citenamefont {Faria}, \citenamefont {Mardegan},\
  and\ \citenamefont {Pagliuso}}]{HoCoGa115}%
  \BibitemOpen
  \bibfield  {author} {\bibinfo {author} {\bibfnamefont {C.}~\bibnamefont
  {Giles}}, \bibinfo {author} {\bibfnamefont {C.}~\bibnamefont {Adriano}},
  \bibinfo {author} {\bibfnamefont {G.~A.}\ \bibnamefont {Faria}}, \bibinfo
  {author} {\bibfnamefont {J.~R.}\ \bibnamefont {Mardegan}}, \ and\ \bibinfo
  {author} {\bibfnamefont {P.~G.}\ \bibnamefont {Pagliuso}},\ }in\ \href@noop
  {} {\emph {\bibinfo {booktitle} {The International Conference on Strongly
  Correlated Electron Systems}}}\ (\bibinfo {year} {2011})\BibitemShut
  {NoStop}%
\bibitem [{\citenamefont {Lora-Serrano}\ \emph
  {et~al.}(2006{\natexlab{b}})\citenamefont {Lora-Serrano}, \citenamefont
  {Ferreira}, \citenamefont {Garcia}, \citenamefont {Miranda}, \citenamefont
  {Giles}, \citenamefont {Duque}, \citenamefont {Granado},\ and\ \citenamefont
  {Pagliuso}}]{TbRh218}%
  \BibitemOpen
  \bibfield  {author} {\bibinfo {author} {\bibfnamefont {R.}~\bibnamefont
  {Lora-Serrano}}, \bibinfo {author} {\bibfnamefont {L.~M.}\ \bibnamefont
  {Ferreira}}, \bibinfo {author} {\bibfnamefont {D.}~\bibnamefont {Garcia}},
  \bibinfo {author} {\bibfnamefont {E.}~\bibnamefont {Miranda}}, \bibinfo
  {author} {\bibfnamefont {C.}~\bibnamefont {Giles}}, \bibinfo {author}
  {\bibfnamefont {J.}~\bibnamefont {Duque}}, \bibinfo {author} {\bibfnamefont
  {E.}~\bibnamefont {Granado}}, \ and\ \bibinfo {author} {\bibfnamefont
  {P.}~\bibnamefont {Pagliuso}},\ }\href {\doibase 10.1016/j.physb.2006.06.035}
  {\bibfield  {journal} {\bibinfo  {journal} {Physica B: Condensed Matter}\
  }\textbf {\bibinfo {volume} {384}},\ \bibinfo {pages} {326 } (\bibinfo {year}
  {2006}{\natexlab{b}})},\ \bibinfo {note} {proceedings of the Seventh Latin
  American Workshop on Magnetism, Magnetic Materials and their
  Applications}\BibitemShut {NoStop}%
\bibitem [{\citenamefont {Granado}\ \emph {et~al.}(2004)\citenamefont
  {Granado}, \citenamefont {Pagliuso}, \citenamefont {Giles}, \citenamefont
  {Lora-Serrano}, \citenamefont {Yokaichiya},\ and\ \citenamefont
  {Sarrao}}]{GdIr218}%
  \BibitemOpen
  \bibfield  {author} {\bibinfo {author} {\bibfnamefont {E.}~\bibnamefont
  {Granado}}, \bibinfo {author} {\bibfnamefont {P.~G.}\ \bibnamefont
  {Pagliuso}}, \bibinfo {author} {\bibfnamefont {C.}~\bibnamefont {Giles}},
  \bibinfo {author} {\bibfnamefont {R.}~\bibnamefont {Lora-Serrano}}, \bibinfo
  {author} {\bibfnamefont {F.}~\bibnamefont {Yokaichiya}}, \ and\ \bibinfo
  {author} {\bibfnamefont {J.~L.}\ \bibnamefont {Sarrao}},\ }\href {\doibase
  10.1103/PhysRevB.69.144411} {\bibfield  {journal} {\bibinfo  {journal} {Phys.
  Rev. B}\ }\textbf {\bibinfo {volume} {69}},\ \bibinfo {pages} {144411}
  (\bibinfo {year} {2004})}\BibitemShut {NoStop}%
\bibitem [{\citenamefont {Adriano}\ \emph {et~al.}(2007)\citenamefont
  {Adriano}, \citenamefont {Lora-Serrano}, \citenamefont {Giles}, \citenamefont
  {de~Bergevin}, \citenamefont {Lang}, \citenamefont {Srajer}, \citenamefont
  {Mazzoli}, \citenamefont {Paolasini},\ and\ \citenamefont
  {Pagliuso}}]{SmIr218}%
  \BibitemOpen
  \bibfield  {author} {\bibinfo {author} {\bibfnamefont {C.}~\bibnamefont
  {Adriano}}, \bibinfo {author} {\bibfnamefont {R.}~\bibnamefont
  {Lora-Serrano}}, \bibinfo {author} {\bibfnamefont {C.}~\bibnamefont {Giles}},
  \bibinfo {author} {\bibfnamefont {F.}~\bibnamefont {de~Bergevin}}, \bibinfo
  {author} {\bibfnamefont {J.~C.}\ \bibnamefont {Lang}}, \bibinfo {author}
  {\bibfnamefont {G.}~\bibnamefont {Srajer}}, \bibinfo {author} {\bibfnamefont
  {C.}~\bibnamefont {Mazzoli}}, \bibinfo {author} {\bibfnamefont
  {L.}~\bibnamefont {Paolasini}}, \ and\ \bibinfo {author} {\bibfnamefont
  {P.~G.}\ \bibnamefont {Pagliuso}},\ }\href {\doibase
  10.1103/PhysRevB.76.104515} {\bibfield  {journal} {\bibinfo  {journal} {Phys.
  Rev. B}\ }\textbf {\bibinfo {volume} {76}},\ \bibinfo {pages} {104515}
  (\bibinfo {year} {2007})}\BibitemShut {NoStop}%
\bibitem [{\citenamefont {Hieu}\ \emph {et~al.}(2005)\citenamefont {Hieu},
  \citenamefont {Shishido}, \citenamefont {Thamizhavel}, \citenamefont
  {Settai}, \citenamefont {Araki}, \citenamefont {Nozue}, \citenamefont
  {Matsuda}, \citenamefont {Haga}, \citenamefont {Takeuchi}, \citenamefont
  {Harima},\ and\ \citenamefont {\'{O}nuki}}]{Pr}%
  \BibitemOpen
  \bibfield  {author} {\bibinfo {author} {\bibfnamefont {N.~V.}\ \bibnamefont
  {Hieu}}, \bibinfo {author} {\bibfnamefont {H.}~\bibnamefont {Shishido}},
  \bibinfo {author} {\bibfnamefont {A.}~\bibnamefont {Thamizhavel}}, \bibinfo
  {author} {\bibfnamefont {R.}~\bibnamefont {Settai}}, \bibinfo {author}
  {\bibfnamefont {S.}~\bibnamefont {Araki}}, \bibinfo {author} {\bibfnamefont
  {Y.}~\bibnamefont {Nozue}}, \bibinfo {author} {\bibfnamefont {T.~D.}\
  \bibnamefont {Matsuda}}, \bibinfo {author} {\bibfnamefont {Y.}~\bibnamefont
  {Haga}}, \bibinfo {author} {\bibfnamefont {T.}~\bibnamefont {Takeuchi}},
  \bibinfo {author} {\bibfnamefont {H.}~\bibnamefont {Harima}}, \ and\ \bibinfo
  {author} {\bibfnamefont {Y.}~\bibnamefont {\'{O}nuki}},\ }\href@noop {}
  {\bibfield  {journal} {\bibinfo  {journal} {Journal of the Physical Society
  of Japan}\ }\textbf {\bibinfo {volume} {74}},\ \bibinfo {pages} {3320}
  (\bibinfo {year} {2005})}\BibitemShut {NoStop}%
\bibitem [{\citenamefont {Duque}\ \emph {et~al.}(2011)\citenamefont {Duque},
  \citenamefont {Serrano}, \citenamefont {Garcia}, \citenamefont {Bufaical},
  \citenamefont {Ferreira}, \citenamefont {Pagliuso},\ and\ \citenamefont
  {Miranda}}]{NdRh115-phase}%
  \BibitemOpen
  \bibfield  {author} {\bibinfo {author} {\bibfnamefont {J.}~\bibnamefont
  {Duque}}, \bibinfo {author} {\bibfnamefont {R.~L.}\ \bibnamefont {Serrano}},
  \bibinfo {author} {\bibfnamefont {D.}~\bibnamefont {Garcia}}, \bibinfo
  {author} {\bibfnamefont {L.}~\bibnamefont {Bufaical}}, \bibinfo {author}
  {\bibfnamefont {L.}~\bibnamefont {Ferreira}}, \bibinfo {author}
  {\bibfnamefont {P.}~\bibnamefont {Pagliuso}}, \ and\ \bibinfo {author}
  {\bibfnamefont {E.}~\bibnamefont {Miranda}},\ }\href {\doibase
  10.1016/j.jmmm.2010.11.077} {\bibfield  {journal} {\bibinfo  {journal}
  {Journal of Magnetism and Magnetic Materials}\ }\textbf {\bibinfo {volume}
  {323}},\ \bibinfo {pages} {954 } (\bibinfo {year} {2011})}\BibitemShut
  {NoStop}%
\bibitem [{\citenamefont {Cermak}\ \emph {et~al.}(2012)\citenamefont {Cermak},
  \citenamefont {Kratochvilova}, \citenamefont {Pajskr},\ and\ \citenamefont
  {Javorsky}}]{RRh218-our}%
  \BibitemOpen
  \bibfield  {author} {\bibinfo {author} {\bibfnamefont {P.}~\bibnamefont
  {Cermak}}, \bibinfo {author} {\bibfnamefont {M.}~\bibnamefont
  {Kratochvilova}}, \bibinfo {author} {\bibfnamefont {K.}~\bibnamefont
  {Pajskr}}, \ and\ \bibinfo {author} {\bibfnamefont {P.}~\bibnamefont
  {Javorsky}},\ }\href {http://stacks.iop.org/0953-8984/24/i=20/a=206005}
  {\bibfield  {journal} {\bibinfo  {journal} {Journal of Physics: Condensed
  Matter}\ }\textbf {\bibinfo {volume} {24}},\ \bibinfo {pages} {206005}
  (\bibinfo {year} {2012})}\BibitemShut {NoStop}%
\bibitem [{\citenamefont {Hieu}\ \emph {et~al.}(2006)\citenamefont {Hieu},
  \citenamefont {Shishido}, \citenamefont {Takeuchi}, \citenamefont
  {Thamizhavel}, \citenamefont {Nakashima}, \citenamefont {Sugiyama},
  \citenamefont {Settai}, \citenamefont {Matsuda}, \citenamefont {Haga},
  \citenamefont {Hagiwara}, \citenamefont {Kindo},\ and\ \citenamefont
  {\={O}nuki}}]{RRh115-CF2006}%
  \BibitemOpen
  \bibfield  {author} {\bibinfo {author} {\bibfnamefont {N.~V.}\ \bibnamefont
  {Hieu}}, \bibinfo {author} {\bibfnamefont {H.}~\bibnamefont {Shishido}},
  \bibinfo {author} {\bibfnamefont {T.}~\bibnamefont {Takeuchi}}, \bibinfo
  {author} {\bibfnamefont {A.}~\bibnamefont {Thamizhavel}}, \bibinfo {author}
  {\bibfnamefont {H.}~\bibnamefont {Nakashima}}, \bibinfo {author}
  {\bibfnamefont {K.}~\bibnamefont {Sugiyama}}, \bibinfo {author}
  {\bibfnamefont {R.}~\bibnamefont {Settai}}, \bibinfo {author} {\bibfnamefont
  {T.~D.}\ \bibnamefont {Matsuda}}, \bibinfo {author} {\bibfnamefont
  {Y.}~\bibnamefont {Haga}}, \bibinfo {author} {\bibfnamefont {M.}~\bibnamefont
  {Hagiwara}}, \bibinfo {author} {\bibfnamefont {K.}~\bibnamefont {Kindo}}, \
  and\ \bibinfo {author} {\bibfnamefont {Y.}~\bibnamefont {\={O}nuki}},\ }\href
  {\doibase 10.1143/JPSJ.75.074708} {\bibfield  {journal} {\bibinfo  {journal}
  {Journal of the Physical Society of Japan}\ }\textbf {\bibinfo {volume}
  {75}},\ \bibinfo {pages} {074708} (\bibinfo {year} {2006})}\BibitemShut
  {NoStop}%
\bibitem [{\citenamefont {Joshi}\ \emph {et~al.}(2008)\citenamefont {Joshi},
  \citenamefont {Nagalakshmi}, \citenamefont {Dhar},\ and\ \citenamefont
  {Thamizhavel}}]{RCoGa218-bulk}%
  \BibitemOpen
  \bibfield  {author} {\bibinfo {author} {\bibfnamefont {D.~A.}\ \bibnamefont
  {Joshi}}, \bibinfo {author} {\bibfnamefont {R.}~\bibnamefont {Nagalakshmi}},
  \bibinfo {author} {\bibfnamefont {S.~K.}\ \bibnamefont {Dhar}}, \ and\
  \bibinfo {author} {\bibfnamefont {A.}~\bibnamefont {Thamizhavel}},\ }\href
  {\doibase 10.1103/PhysRevB.77.174420} {\bibfield  {journal} {\bibinfo
  {journal} {Phys. Rev. B}\ }\textbf {\bibinfo {volume} {77}},\ \bibinfo
  {pages} {174420} (\bibinfo {year} {2008})}\BibitemShut {NoStop}%
\bibitem [{\citenamefont {Hieu}\ \emph {et~al.}(2007)\citenamefont {Hieu},
  \citenamefont {Takeuchi}, \citenamefont {Shishido}, \citenamefont {Tonohiro},
  \citenamefont {Yamada}, \citenamefont {Nakashima}, \citenamefont {Sugiyama},
  \citenamefont {Settai}, \citenamefont {Matsuda}, \citenamefont {Haga},
  \citenamefont {Hagiwara}, \citenamefont {Kindo}, \citenamefont {Araki},
  \citenamefont {Nozue},\ and\ \citenamefont {\={O}nuki}}]{RRh115-CF2007}%
  \BibitemOpen
  \bibfield  {author} {\bibinfo {author} {\bibfnamefont {N.~V.}\ \bibnamefont
  {Hieu}}, \bibinfo {author} {\bibfnamefont {T.}~\bibnamefont {Takeuchi}},
  \bibinfo {author} {\bibfnamefont {H.}~\bibnamefont {Shishido}}, \bibinfo
  {author} {\bibfnamefont {C.}~\bibnamefont {Tonohiro}}, \bibinfo {author}
  {\bibfnamefont {T.}~\bibnamefont {Yamada}}, \bibinfo {author} {\bibfnamefont
  {H.}~\bibnamefont {Nakashima}}, \bibinfo {author} {\bibfnamefont
  {K.}~\bibnamefont {Sugiyama}}, \bibinfo {author} {\bibfnamefont
  {R.}~\bibnamefont {Settai}}, \bibinfo {author} {\bibfnamefont {T.~D.}\
  \bibnamefont {Matsuda}}, \bibinfo {author} {\bibfnamefont {Y.}~\bibnamefont
  {Haga}}, \bibinfo {author} {\bibfnamefont {M.}~\bibnamefont {Hagiwara}},
  \bibinfo {author} {\bibfnamefont {K.}~\bibnamefont {Kindo}}, \bibinfo
  {author} {\bibfnamefont {S.}~\bibnamefont {Araki}}, \bibinfo {author}
  {\bibfnamefont {Y.}~\bibnamefont {Nozue}}, \ and\ \bibinfo {author}
  {\bibfnamefont {Y.}~\bibnamefont {\={O}nuki}},\ }\href {\doibase
  10.1143/JPSJ.76.064702} {\bibfield  {journal} {\bibinfo  {journal} {Journal
  of the Physical Society of Japan}\ }\textbf {\bibinfo {volume} {76}},\
  \bibinfo {pages} {064702} (\bibinfo {year} {2007})}\BibitemShut {NoStop}%
\bibitem [{\citenamefont {Uhl\'{i}\v{r}ov\'{a}}\ and\ \citenamefont
  {Sechovsk\'{y}}(2009)}]{growth}%
  \BibitemOpen
  \bibfield  {author} {\bibinfo {author} {\bibfnamefont {K.}~\bibnamefont
  {Uhl\'{i}\v{r}ov\'{a}}}\ and\ \bibinfo {author} {\bibfnamefont
  {V.}~\bibnamefont {Sechovsk\'{y}}},\ }\href@noop {} {\bibfield  {journal}
  {\bibinfo  {journal} {Int. J. Mater. Res.}\ }\textbf {\bibinfo {volume}
  {100}},\ \bibinfo {pages} {1242} (\bibinfo {year} {2009})}\BibitemShut
  {NoStop}%
\bibitem [{\citenamefont {McIntyre}\ \emph {et~al.}(2006)\citenamefont
  {McIntyre}, \citenamefont {Lem\'{e}e-Cailleau},\ and\ \citenamefont
  {Wilkinson}}]{VIVALDI}%
  \BibitemOpen
  \bibfield  {author} {\bibinfo {author} {\bibfnamefont {G.~J.}\ \bibnamefont
  {McIntyre}}, \bibinfo {author} {\bibfnamefont {M.-H.}\ \bibnamefont
  {Lem\'{e}e-Cailleau}}, \ and\ \bibinfo {author} {\bibfnamefont
  {C.}~\bibnamefont {Wilkinson}},\ }\href@noop {} {\bibfield  {journal}
  {\bibinfo  {journal} {Physica B}\ }\textbf {\bibinfo {volume} {385-386}},\
  \bibinfo {pages} {1055} (\bibinfo {year} {2006})}\BibitemShut {NoStop}%
\bibitem [{\citenamefont {Ouladdiaf}\ \emph {et~al.}(2011)\citenamefont
  {Ouladdiaf}, \citenamefont {Archer}, \citenamefont {Allibon}, \citenamefont
  {Decarpentrie}, \citenamefont {Lem{\'{e}}e-Cailleau}, \citenamefont
  {Rodr{\'\i}guez-Carvajal}, \citenamefont {Hewat}, \citenamefont {York},
  \citenamefont {Brau},\ and\ \citenamefont {McIntyre}}]{CYCLOPS}%
  \BibitemOpen
  \bibfield  {author} {\bibinfo {author} {\bibfnamefont {B.}~\bibnamefont
  {Ouladdiaf}}, \bibinfo {author} {\bibfnamefont {J.}~\bibnamefont {Archer}},
  \bibinfo {author} {\bibfnamefont {J.~R.}\ \bibnamefont {Allibon}}, \bibinfo
  {author} {\bibfnamefont {P.}~\bibnamefont {Decarpentrie}}, \bibinfo {author}
  {\bibfnamefont {M.-H.}\ \bibnamefont {Lem{\'{e}}e-Cailleau}}, \bibinfo
  {author} {\bibfnamefont {J.}~\bibnamefont {Rodr{\'\i}guez-Carvajal}},
  \bibinfo {author} {\bibfnamefont {A.~W.}\ \bibnamefont {Hewat}}, \bibinfo
  {author} {\bibfnamefont {S.}~\bibnamefont {York}}, \bibinfo {author}
  {\bibfnamefont {D.}~\bibnamefont {Brau}}, \ and\ \bibinfo {author}
  {\bibfnamefont {G.~J.}\ \bibnamefont {McIntyre}},\ }\href {\doibase
  10.1107/S0021889811006765} {\bibfield  {journal} {\bibinfo  {journal}
  {Journal of Applied Crystallography}\ }\textbf {\bibinfo {volume} {44}},\
  \bibinfo {pages} {392} (\bibinfo {year} {2011})}\BibitemShut {NoStop}%
\bibitem [{\citenamefont {Rodriguez-Carvajal}\ \emph {et~al.}()\citenamefont
  {Rodriguez-Carvajal}, \citenamefont {Fuentes-Montero},\ and\ \citenamefont
  {\v{C}erm\'{a}k}}]{ESMERALDA}%
  \BibitemOpen
  \bibfield  {author} {\bibinfo {author} {\bibfnamefont {J.}~\bibnamefont
  {Rodriguez-Carvajal}}, \bibinfo {author} {\bibfnamefont {L.}~\bibnamefont
  {Fuentes-Montero}}, \ and\ \bibinfo {author} {\bibfnamefont {P.}~\bibnamefont
  {\v{C}erm\'{a}k}},\ }\href@noop {} {\bibinfo  {journal} {to be published. See
  the repository https://forge.ill.eu/projects/sxtalsoft/}\ }\BibitemShut
  {NoStop}%
\bibitem [{\citenamefont {Filhol}()}]{rafd9}%
  \BibitemOpen
\bibfield  {journal} {  }\bibfield  {author} {\bibinfo {author} {\bibfnamefont
  {A.}~\bibnamefont {Filhol}},\ }\href@noop {} {\enquote {\bibinfo {title}
  {Program for the refinement of the ub-matrix, wavelength, cell parameters and
  instrument zero-shifts in a single crystal diffraction experiment},}\
  }\bibinfo {note} {See web page
  https://forge.epn-campus.eu/projects/sxtalsoft/repository/show/rafub}\BibitemShut
  {NoStop}%
\bibitem [{\citenamefont {Wilkinson}\ \emph {et~al.}(1988)\citenamefont
  {Wilkinson}, \citenamefont {Khamis}, \citenamefont {Stansfield},\ and\
  \citenamefont {McIntyre}}]{racer}%
  \BibitemOpen
  \bibfield  {author} {\bibinfo {author} {\bibfnamefont {C.}~\bibnamefont
  {Wilkinson}}, \bibinfo {author} {\bibfnamefont {H.}~\bibnamefont {Khamis}},
  \bibinfo {author} {\bibfnamefont {R.}~\bibnamefont {Stansfield}}, \ and\
  \bibinfo {author} {\bibfnamefont {G.}~\bibnamefont {McIntyre}},\ }\href@noop
  {} {\bibfield  {journal} {\bibinfo  {journal} {Journal of Applied
  Crystallography}\ }\textbf {\bibinfo {volume} {21}},\ \bibinfo {pages} {471}
  (\bibinfo {year} {1988})},\ \bibinfo {note} {see web page:
  https://forge.epn-campus.eu/projects/sxtalsoft/repository/show/racer}\BibitemShut
  {NoStop}%
\bibitem [{\citenamefont {Coppens}()}]{datap}%
  \BibitemOpen
  \bibfield  {author} {\bibinfo {author} {\bibfnamefont {P.}~\bibnamefont
  {Coppens}},\ }\href@noop {} {\enquote {\bibinfo {title} {Absorbtion
  correction program datap},}\ }\bibinfo {note} {See web page
  https://forge.epn-campus.eu/projects/sxtalsoft/wiki/Datap}\BibitemShut
  {NoStop}%
\bibitem [{\citenamefont {Rodriguez-Carvajal}(1993)}]{Fullprof}%
  \BibitemOpen
  \bibfield  {author} {\bibinfo {author} {\bibfnamefont {J.}~\bibnamefont
  {Rodriguez-Carvajal}},\ }\href@noop {} {\bibfield  {journal} {\bibinfo
  {journal} {Physica B}\ }\textbf {\bibinfo {volume} {192}},\ \bibinfo {pages}
  {55} (\bibinfo {year} {1993})}\BibitemShut {NoStop}%
\bibitem [{\citenamefont {Zachariesen}(1967)}]{extinkce}%
  \BibitemOpen
  \bibfield  {author} {\bibinfo {author} {\bibfnamefont {W.~H.}\ \bibnamefont
  {Zachariesen}},\ }\href@noop {} {\bibfield  {journal} {\bibinfo  {journal}
  {Acta Cryst.}\ }\textbf {\bibinfo {volume} {23}},\ \bibinfo {pages} {558}
  (\bibinfo {year} {1967})}\BibitemShut {NoStop}%
\bibitem [{\citenamefont {\v{C}erm\'{a}k}\ \emph {et~al.}(2013)\citenamefont
  {\v{C}erm\'{a}k}, \citenamefont {Divi\v{s}}, \citenamefont
  {Kratochv\'{i}lov\'{a}},\ and\ \citenamefont {Javorsk\'{y}}}]{nonmag}%
  \BibitemOpen
  \bibfield  {author} {\bibinfo {author} {\bibfnamefont {P.}~\bibnamefont
  {\v{C}erm\'{a}k}}, \bibinfo {author} {\bibfnamefont {M.}~\bibnamefont
  {Divi\v{s}}}, \bibinfo {author} {\bibfnamefont {M.}~\bibnamefont
  {Kratochv\'{i}lov\'{a}}}, \ and\ \bibinfo {author} {\bibfnamefont
  {P.}~\bibnamefont {Javorsk\'{y}}},\ }\href@noop {} {\bibfield  {journal}
  {\bibinfo  {journal} {Solid State Communications}\ }\textbf {\bibinfo
  {volume} {163}},\ \bibinfo {pages} {55} (\bibinfo {year} {2013})}\BibitemShut
  {NoStop}%
\bibitem [{\citenamefont {Pagliuso}\ \emph {et~al.}(2000)\citenamefont
  {Pagliuso}, \citenamefont {Thompson}, \citenamefont {Hundley},\ and\
  \citenamefont {Sarrao}}]{Pagliuso2000}%
  \BibitemOpen
  \bibfield  {author} {\bibinfo {author} {\bibfnamefont {P.~G.}\ \bibnamefont
  {Pagliuso}}, \bibinfo {author} {\bibfnamefont {J.~D.}\ \bibnamefont
  {Thompson}}, \bibinfo {author} {\bibfnamefont {M.~F.}\ \bibnamefont
  {Hundley}}, \ and\ \bibinfo {author} {\bibfnamefont {J.~L.}\ \bibnamefont
  {Sarrao}},\ }\href@noop {} {\bibfield  {journal} {\bibinfo  {journal} {Phys.
  Rev. B}\ }\textbf {\bibinfo {volume} {62}},\ \bibinfo {pages} {12266}
  (\bibinfo {year} {2000})}\BibitemShut {NoStop}%
\bibitem [{\citenamefont {Bertaut}(1968)}]{RepAnaBertaut}%
  \BibitemOpen
  \bibfield  {author} {\bibinfo {author} {\bibfnamefont {E.~F.}\ \bibnamefont
  {Bertaut}},\ }\href {\doibase 10.1107/S0567739468000306} {\bibfield
  {journal} {\bibinfo  {journal} {Acta Crystallographica Section A}\ }\textbf
  {\bibinfo {volume} {24}},\ \bibinfo {pages} {217} (\bibinfo {year}
  {1968})}\BibitemShut {NoStop}%
\bibitem [{\citenamefont {Aroyo}\ \emph
  {et~al.}(2006{\natexlab{a}})\citenamefont {Aroyo}, \citenamefont
  {Perez-Mato}, \citenamefont {Capillas}, \citenamefont {Kroumova},
  \citenamefont {Ivantchev}, \citenamefont {Madariaga}, \citenamefont {Kirov},\
  and\ \citenamefont {Wondratschek}}]{BCS1}%
  \BibitemOpen
  \bibfield  {author} {\bibinfo {author} {\bibfnamefont {M.~I.}\ \bibnamefont
  {Aroyo}}, \bibinfo {author} {\bibfnamefont {J.~M.}\ \bibnamefont
  {Perez-Mato}}, \bibinfo {author} {\bibfnamefont {C.}~\bibnamefont
  {Capillas}}, \bibinfo {author} {\bibfnamefont {E.}~\bibnamefont {Kroumova}},
  \bibinfo {author} {\bibfnamefont {S.}~\bibnamefont {Ivantchev}}, \bibinfo
  {author} {\bibfnamefont {G.}~\bibnamefont {Madariaga}}, \bibinfo {author}
  {\bibfnamefont {A.}~\bibnamefont {Kirov}}, \ and\ \bibinfo {author}
  {\bibfnamefont {H.}~\bibnamefont {Wondratschek}},\ }\href {\doibase
  10.1524/zkri.2006.221.1.15} {\bibfield  {journal} {\bibinfo  {journal} {Z.
  Krist.}\ }\textbf {\bibinfo {volume} {221}},\ \bibinfo {pages} {15} (\bibinfo
  {year} {2006}{\natexlab{a}})},\ \bibinfo {note} {see web page
  http://www.cryst.ehu.es}\BibitemShut {NoStop}%
\bibitem [{\citenamefont {Aroyo}\ \emph
  {et~al.}(2006{\natexlab{b}})\citenamefont {Aroyo}, \citenamefont {Kirov},
  \citenamefont {Capillas}, \citenamefont {Perez-Mato},\ and\ \citenamefont
  {Wondratschek}}]{BCS2}%
  \BibitemOpen
  \bibfield  {author} {\bibinfo {author} {\bibfnamefont {M.~I.}\ \bibnamefont
  {Aroyo}}, \bibinfo {author} {\bibfnamefont {A.}~\bibnamefont {Kirov}},
  \bibinfo {author} {\bibfnamefont {C.}~\bibnamefont {Capillas}}, \bibinfo
  {author} {\bibfnamefont {J.~M.}\ \bibnamefont {Perez-Mato}}, \ and\ \bibinfo
  {author} {\bibfnamefont {H.}~\bibnamefont {Wondratschek}},\ }\href {\doibase
  10.1107/S0108767305040286} {\bibfield  {journal} {\bibinfo  {journal} {Acta
  Cryst. A}\ }\textbf {\bibinfo {volume} {62}},\ \bibinfo {pages} {115}
  (\bibinfo {year} {2006}{\natexlab{b}})}\BibitemShut {NoStop}%
\bibitem [{\citenamefont {Stokes}\ \emph {et~al.}()\citenamefont {Stokes},
  \citenamefont {Hatch},\ and\ \citenamefont {Campbell}}]{ISOTROPY}%
  \BibitemOpen
  \bibfield  {author} {\bibinfo {author} {\bibfnamefont {H.~T.}\ \bibnamefont
  {Stokes}}, \bibinfo {author} {\bibfnamefont {D.~M.}\ \bibnamefont {Hatch}}, \
  and\ \bibinfo {author} {\bibfnamefont {B.~J.}\ \bibnamefont {Campbell}},\
  }\href@noop {} {\enquote {\bibinfo {title} {Isotropy software suite,
  iso.byu.edu.}}\ }\BibitemShut {NoStop}%
\bibitem [{\citenamefont {Javorsk\'y}\ \emph {et~al.}(2014)\citenamefont
  {Javorsk\'y}, \citenamefont {Pajskr}, \citenamefont {Klicpera}, \citenamefont
  {\v{C}erm\'{a}k}, \citenamefont {Skourski},\ and\ \citenamefont
  {Andreev}}]{TbNd-high-field}%
  \BibitemOpen
  \bibfield  {author} {\bibinfo {author} {\bibfnamefont {P.}~\bibnamefont
  {Javorsk\'y}}, \bibinfo {author} {\bibfnamefont {K.}~\bibnamefont {Pajskr}},
  \bibinfo {author} {\bibfnamefont {M.}~\bibnamefont {Klicpera}}, \bibinfo
  {author} {\bibfnamefont {P.}~\bibnamefont {\v{C}erm\'{a}k}}, \bibinfo
  {author} {\bibfnamefont {Y.}~\bibnamefont {Skourski}}, \ and\ \bibinfo
  {author} {\bibfnamefont {A.}~\bibnamefont {Andreev}},\ }\href {\doibase
  http://dx.doi.org/10.1016/j.jallcom.2014.02.042} {\bibfield  {journal}
  {\bibinfo  {journal} {Journal of Alloys and Compounds}\ }\textbf {\bibinfo
  {volume} {598}},\ \bibinfo {pages} {278 } (\bibinfo {year}
  {2014})}\BibitemShut {NoStop}%
\bibitem [{\citenamefont {Szytula}\ and\ \citenamefont
  {Leciejewicz}(1994)}]{szytula1994handbook}%
  \BibitemOpen
  \bibfield  {author} {\bibinfo {author} {\bibfnamefont {A.}~\bibnamefont
  {Szytula}}\ and\ \bibinfo {author} {\bibfnamefont {J.}~\bibnamefont
  {Leciejewicz}},\ }\href {http://books.google.de/books?id=-tgM8oAQcdcC} {\emph
  {\bibinfo {title} {Handbook of Crystal Structures and Magnetic Properties of
  Rare Earth Intermetallics}}}\ (\bibinfo  {publisher} {Taylor \& Francis},\
  \bibinfo {year} {1994})\BibitemShut {NoStop}%
\bibitem [{\citenamefont {Iwata}\ \emph {et~al.}(1990)\citenamefont {Iwata},
  \citenamefont {Honda}, \citenamefont {Shigeoka}, \citenamefont {Hashimoto},\
  and\ \citenamefont {Fujii}}]{DyCo2Si2-Iwata1990}%
  \BibitemOpen
  \bibfield  {author} {\bibinfo {author} {\bibfnamefont {N.}~\bibnamefont
  {Iwata}}, \bibinfo {author} {\bibfnamefont {K.}~\bibnamefont {Honda}},
  \bibinfo {author} {\bibfnamefont {T.}~\bibnamefont {Shigeoka}}, \bibinfo
  {author} {\bibfnamefont {Y.}~\bibnamefont {Hashimoto}}, \ and\ \bibinfo
  {author} {\bibfnamefont {H.}~\bibnamefont {Fujii}},\ }\href {\doibase
  http://dx.doi.org/10.1016/S0304-8853(10)80022-8} {\bibfield  {journal}
  {\bibinfo  {journal} {Journal of Magnetism and Magnetic Materials}\ }\textbf
  {\bibinfo {volume} {90–91}},\ \bibinfo {pages} {63 } (\bibinfo {year}
  {1990})}\BibitemShut {NoStop}%
\bibitem [{\citenamefont {Izyumov}\ \emph {et~al.}(1979)\citenamefont
  {Izyumov}, \citenamefont {Naish},\ and\ \citenamefont
  {Petrov}}]{Exchange_Multiplets}%
  \BibitemOpen
  \bibfield  {author} {\bibinfo {author} {\bibfnamefont {Y.}~\bibnamefont
  {Izyumov}}, \bibinfo {author} {\bibfnamefont {V.}~\bibnamefont {Naish}}, \
  and\ \bibinfo {author} {\bibfnamefont {S.}~\bibnamefont {Petrov}},\
  }\href@noop {} {\bibfield  {journal} {\bibinfo  {journal} {Journal or
  Magnetism and Magnetic Materials}\ }\textbf {\bibinfo {volume} {13}},\
  \bibinfo {pages} {275} (\bibinfo {year} {1979})}\BibitemShut {NoStop}%
\end{thebibliography}%

\end{document}